\documentclass[pra,eqsecnum,amsmath,amssymb,dvips,twocolumn,showpacs]{revtex4}
\usepackage{epsfig}
\usepackage{bm}
\usepackage{subfigure}

\newcommand{\beq}{\begin{equation}}
\newcommand{\eeq}{\end{equation}}
\newcommand{\bea}{\begin{eqnarray}}
\newcommand{\eea}{\end{eqnarray}}

\begin{document}

\title{Lattice field theory simulations of graphene}

\author{Joaqu\'{\i}n E. Drut$^1$ and Timo A. L\"ahde$^2$ }
\affiliation{$^1$Department of Physics, The Ohio State University, Columbus, OH 43210--1117, USA}
\affiliation{$^2$Department of Physics, University of Washington, Seattle, WA 98195--1560, USA}

\date {\today}

\begin{abstract}
We discuss the Monte Carlo method of simulating lattice field theories as a means of studying the low-energy effective 
theory of graphene. We also report on simulational results obtained using the Metropolis and Hybrid Monte Carlo methods for 
the chiral condensate, which is the order parameter for the semimetal-insulator transition in graphene, induced by the 
Coulomb interaction between the massless electronic quasiparticles. The critical coupling and the associated exponents of 
this transition are determined by means of the logarithmic derivative of the chiral condensate and an equation-of-state 
analysis. A thorough discussion of finite-size effects is given, along with several tests of our calculational framework. 
These results strengthen the case for an insulating phase in suspended graphene, and indicate that the semimetal-insulator 
transition is likely to be of second order, though exhibiting neither classical critical exponents, nor the predicted 
phenomenon of Miransky scaling.
\end{abstract}


\maketitle


\section{Introduction}
\label{sec:intro}

The recent experimental isolation of single atomic layers of graphite, known as graphene, has provided physicists with a 
novel opportunity to study a strongly coupled system with remarkable many-body and electronic properties, which at the same 
time can be easily manipulated experimentally~\cite{GeimNovoselov,CastroNetoetal}. Even more recently, the advent of 
experiments utilizing samples of suspended graphene, free from the interference of an underlying substrate~\cite{SuspExp}, 
has provided unprecedented insight into the intrinsic properties of graphene. Among other remarkable discoveries, suspended 
graphene has been shown to possess a very high carrier mobility even at room temperature, as well as a markedly 
non-metallic behavior of the conductivity at low temperatures.

\begin{figure}[b]
\subfigure[]
{
\label{fig:Dirac}
\includegraphics[width=.28\columnwidth,viewport=1mm 5mm 35mm 73mm,clip]{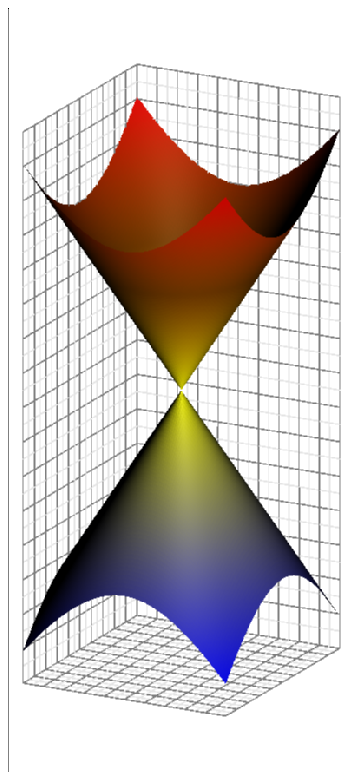}
}
\hspace{.4cm}
\subfigure[]
{
\label{fig:Lattice}
\includegraphics[width=.6\columnwidth]{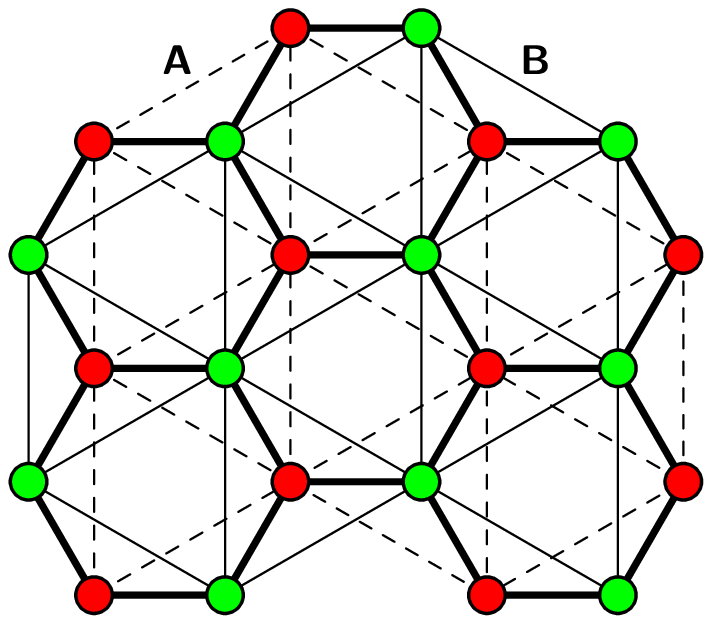}
}
\caption{(Color online)
a) Dirac cone, joining the upper (red) conduction band and the lower (blue) valence band.
b) The hexagonal arrangement of carbon atoms in graphene, with sublattices A~(red dots, thin dashed lines) and 
B~(green dots, thin solid lines).}
\end{figure}

A central property of graphene is that the low-energy electronic spectrum can be described in terms of two flavors of 
massless, four-component fermionic quasiparticles with linear dispersion~\cite{Semenoff}. Indeed, due to the hexagonal 
honeycomb arrangement of the carbon atoms in the graphene lattice, the band structure of graphene exhibits two inequivalent 
(but degenerate) ``Dirac cones'' where the conduction and valence bands touch, as illustrated in 
Fig.~\ref{fig:Dirac}. Since the energy-momentum relation around a Dirac point is linear as in relativistic 
theories, the low-energy description of graphene bears a certain resemblance to massless Quantum Electrodynamics (QED). 
Nevertheless, an important difference is that the Fermi velocity of the quasiparticles in graphene is as low as $v\simeq 
c/300$, whereby the electromagnetic interaction is rendered essentially instantaneous.

Such a description is well-known to account for the physics of graphene on a substrate, where the system exhibits 
semimetallic properties due to the absence of a gap in the electronic spectrum. While suspended graphene has recently 
come under intense experimental investigation~\cite{SuspExp}, its spectrum is yet to be computed in a 
controlled fashion. From the theoretical perspective, the challenging feature of suspended graphene lies in the fact that 
the Coulomb interaction between the quasiparticles is unscreened which, in conjunction with the small Fermi velocity, 
results in a graphene analogue of the fine-structure constant $\alpha_g^{} \gtrsim 1$. At such strong coupling, a dynamical 
transition into a phase fundamentally different from the weakly-coupled semimetallic phase of graphene is a strong 
possibility. In graphene sheets deposited on a substrate, such a transition is effectively inhibited due to the screening 
of the Coulomb interaction by the dielectric.

Our recent work in Ref.~\cite{DrutLahde1} has demonstrated that graphene is expected to undergo a semimetal-insulator 
transition when the substrate is removed. More specifically, evidence was found that the low-energy effective theory of 
graphene undergoes a phase transition involving spontaneous chiral symmetry breaking, which takes place at a critical 
coupling of $\beta_c = 0.072 \pm 0.005$, and within the accuracy of that work the transition appeared to be consistent with 
classical mean-field exponents. The results reported in Ref.~\cite{DrutLahde1} are based on the numerical Monte Carlo 
simulation of a discretized lattice formulation of the low-energy effective theory of graphene, and the calculation of the 
chiral condensate, which is the order parameter for excitonic gap formation. In one possible realization of such an 
insulating state, the equivalence of the triangular sublattices A~and~B, shown in Fig.~\ref{fig:Lattice}, is broken by the 
accumulation of charge carriers of opposite sign on the respective sublattices.

Our results should be compared with those of Refs.~\cite{Leal:2003sg,Gorbar} which are based on a gap equation, where a 
semimetal-insulator transition was found at critical couplings of $\beta_c^{} \sim 0.06$ and $\beta_c^{} \sim 0.03$ 
respectively. While the result of Ref.~\cite{Leal:2003sg} is within the physical range of Coulomb couplings, that of 
Ref.~\cite{Gorbar} is slightly above the largest conceivable value of $\alpha_g^{} \sim 2.16$, which corresponds to graphene 
in vacuum. On the other hand, Refs.~\cite{Gonzalez,Herbut} employed an expansion 
in the inverse number of fermion flavors $N_f^{}$, and found that at large $N_f^{}$, the Coulomb interaction between the 
quasiparticles becomes irrelevant and therefore unable to induce a gap in the electronic spectrum.

In this paper, we explain the details of our Lattice Monte Carlo method, which to our knowledge has not been applied to the 
low-energy theory of graphene (however, Ref.~\cite{HandsStrouthos} has considered a theory related to the strong-coupling 
limit). We also present new calculations supporting the conclusions of Ref.~\cite{DrutLahde1}, but extending the previous 
data set to much larger lattices. In Section~\ref{sec:EffectiveTheory}, we discuss the low-energy effective theory of 
graphene, the corresponding partition function, and the computation of observables upon integration of the fermionic 
degrees of freedom. In Section~\ref{sec:lattice}, we describe the discretization of the effective theory and introduce a 
lattice formulation that respects gauge invariance and avoids the fermion doubling problem while maintaining a certain 
degree of chiral symmetry at finite lattice spacing. In Section~\ref{sec:results}, the results of our simulations are 
presented, with emphasis on the chiral condensate and susceptibility, including a determination of the critical coupling for 
the semimetal-insulator phase transition, and the consequences of our results for the corresponding critical exponents. In 
Section~\ref{sec:tests}, we outline the various tests and cross-checks we have performed in order to validate our results. 
In Section~\ref{sec:discussion} we discuss the possibility of observing the transition experimentally. Finally, in 
Section~\ref{sec:conclusions} we summarize our findings and present a case for continued study.


\section{Low-energy effective theory\label{sec:EffectiveTheory}}

The electronic band structure of graphene close to the Fermi level forms the basis of the low-energy effective theory of 
graphene. This band structure is a reflection of the hexagonal arrangement of the carbon atoms as shown in 
Fig.~\ref{fig:Lattice}, and can be well described by a tight-binding model of the form
\bea
H &=& -t \!\!\!\sum_{\langle i,j \rangle,\sigma=\uparrow,\downarrow}\!\!\!
\left(a_{\sigma,i}^{\dagger} b_{\sigma,j}^{} + \text{H.c.}\right) \nonumber \\
&&-t' \!\!\!\!\!\!\sum_{\langle \langle i,j \rangle \rangle,\sigma=\uparrow,\downarrow}\!\!\!\!\!\!
\left(a_{\sigma,i}^{\dagger} a_{\sigma,j}^{} +  b_{\sigma,i}^{\dagger} b_{\sigma,j}^{} + \text{H.c.}\right), \quad\quad
\eea
as first done by Wallace in Ref.~\cite{Wallace}. The operators $a_{\sigma,i}^\dagger (a_{\sigma,i}^{})$ and 
$b_{\sigma,i}^\dagger (b_{\sigma,i}^{})$ create (annihilate) an electron of spin $\sigma$ at location $i$ on the A and B 
sublattices, respectively~[see Fig.~\ref{fig:Lattice}]. The first term (involving $t$) takes into account 
nearest-neighbor interactions, and the second term (involving $t'$) the next-to-nearest neighbor ones. Both terms account for 
all spin states. The hopping parameters that give an optimal fit to the experimentally determined band structure of graphene 
are $t \simeq 2.8$~eV and $t' \simeq 0.1$~eV~\cite{Reich}. Third-nearest neighbors have also been considered in 
Ref.~\cite{Reich}, yielding an additional hopping amplitude of $t'' \simeq 0.07$~eV.

We shall follow a somewhat different route based on an Effective Field Theory~(EFT) treatment of 
graphene~\cite{Son,Herbut}, which has the advantage of describing the physics of graphene directly in terms of 
the relevant low-energy degrees of freedom, namely charged massless fermionic quasiparticles. 
The EFT description of graphene has an additional advantage as it allows for the direct study of effects due to the 
unscreened, long-range Coulomb interactions between the quasiparticles. In what follows, we shall formulate a {\it 
continuum} Lagrangian field theory that should be thought of as valid only at low momenta, much smaller than the inverse 
of the interatomic distance in graphene, which is $\sim 1.42$~\AA.


\subsection{Continuum formulation}
\label{subsec:Continuum}

In the EFT framework, graphene is described by a theory of $N_f^{}$ Dirac flavors interacting via an instantaneous Coulomb 
interaction. The action (in Euclidean spacetime) of this theory is
\begin{eqnarray}
S_E^{} &=& -\sum_{a=1}^{N_f^{}} \int d^2x\,dt \: \bar\psi_a^{} \:D[A_0^{}]\: \psi_a^{} 
\nonumber \\
&& +\,\frac{1}{2g^2} \int d^3x\,dt \: (\partial_i^{} A_0^{})^2,
\label{SE}
\end{eqnarray}
where $N_f^{} = 2$ for graphene monolayers, $g^2 = e^2/\epsilon_0^{}$ for graphene in vacuum (suspended graphene), 
$\psi_a^{}$ is a four-component Dirac field in 2+1~dimensions, $A_0^{}$ is a Coulomb field in 3+1 dimensions, and
\begin{eqnarray}
D[A_0^{}] &=& \gamma_0^{} (\partial_0^{} + iA_0^{}) + v\gamma_i^{} \partial_i^{}, \quad i=1,2
\end{eqnarray}
where the Dirac matrices $\gamma_\mu^{}$ satisfy the Euclidean Clifford algebra $\{ \gamma_\mu^{}, \gamma_\nu^{}\} = 
2\delta_{\mu\nu}^{}$. 
The four-component spinor structure accounts for quasiparticle excitations of sublattices~A and~B around the two Dirac
points in the band structure~\cite{Semenoff,Herbut}. The two Dirac points are identified with the two inequivalent 
representations (with opposite parity) of the Dirac matrices in 2+1 dimensions. In graphene monolayers, $N_f^{} = 2$ owing 
to electronic spin, while $N_f^{} = 4$ is related to the case of two decoupled graphene layers, interacting solely via the 
Coulomb interaction. Consideration of arbitrary $N_f^{}$ is also useful, given that an analytic 
treatment~\cite{Gonzalez} is possible in the limit $N_f^{} \to \infty$.

The strength of the Coulomb interaction is controlled by $\alpha_g^{} = e^2/(4 \pi v 
\epsilon_0^{})$, which is the graphene analogue of the fine-structure constant $\alpha \simeq 1/137$ of QED. It is 
straightforward to show that $\alpha_g^{}$ is the only parameter, by rescaling according to
\bea
t' &=& vt, \nonumber \\ 
A_0' &=& A_0^{}/v. 
\label{resc}
\eea
The action~(\ref{SE}) is invariant under spatially uniform gauge transformations (see Sec. \ref{subsec:gauge}). Notice that 
since the $A_0^{}$ field is 3+1 dimensional, one recovers the four-fermion Coulomb interaction
\beq
\frac{
\bar\psi_a^{}(x) \gamma_0^{} \psi_a^{}(x) \:
\bar\psi_b^{}(x') \gamma_0^{} \psi_b^{}(x')}
{|{\bf x} - {\bf x'}|}
\eeq
by integrating out $A_0^{}$. Nevertheless, for our purposes the original form of the action (quadratic in the fermions) as 
given in Eq.(\ref{SE}) is preferable.

A central property of the low-energy EFT is that Eq.~(\ref{SE}) respects a global U$(2 N_f^{})$ chiral 
symmetry under the transformations  
\bea
\psi_a^{} &\rightarrow& \exp(i \Gamma_j^{} \alpha_j^{})\,\psi_a^{}
\eea
where the matrices $\Gamma_j^{}$ are the $(2 N_f^{})^2$ Hermitian generators of U$(2 N_f^{})$, such that for the 
case of graphene monolayers, the group is U$(4)$. The generators can be constructed by first choosing a representation
for the $\gamma_\mu^{}$, such as
\beq
\label{gammamatrices}
 \gamma_0^{} = \left(\begin{array}{cc} 
    \sigma_3^{} & 0 \\ 0 & -\sigma_3^{} \end{array}\right),
 \qquad 
 \gamma_i^{} = \left(\begin{array}{cc} 
    \sigma_i^{} & 0 \\ 0 & -\sigma_i^{} \end{array}\right)
\eeq
where the $\sigma_i^{}$ are Pauli matrices. Adding the identity to this set yields the generators of U$(2)$, since they form 
a set of four linearly independent Hermitian matrices. It should be noted that the choice of any particular representation 
for the $\gamma_\mu^{}$ is completely arbitrary and is not necessary for any calculational purpose, as all relevant 
information is provided by the Clifford algebra. However, the identification of the spinor degrees of freedom with any 
particular Dirac point and graphene sublattice is dependent on the chosen representation. 

In order to arrive at the 
generators of U$(4)$, one can take the direct product of each of the abovementioned generators of U$(2)$ by 
$\{\openone, \sigma^1,\sigma^2, \sigma^3 \}$, where the latter operate in flavor space. In this way, one obtains a set of 
precisely sixteen linearly independent Hermitian matrices, forming the generators of U$(4)$. Significantly, this chiral 
symmetry can be spontaneously broken down to \mbox{U$(2)\times$U$(2)$}, in which case the excitonic condensate $\langle \bar 
\psi \psi\rangle$ acquires a non-vanishing value, signaling the formation of quasiparticle-hole bound states. The same group 
structure is obtained by adding to Eq.~(\ref{SE}) a parity invariant (Dirac) mass term
\beq
\label{eq:massterm}
\int d^2x\,dt \: m_0^{} \bar\psi_a^{} \psi_a^{},
\eeq
which breaks chiral symmetry explicitly. The remaining unbroken generators are then $\{\openone, \sigma^3 \}$, which 
correspond to uniform phase rotations of both flavors with the same phase, and with equal and opposite phases, respectively. 
For the extended theory with $N_f^{}$ flavors, the symmetry breaking pattern is \mbox{U$(2 N_f^{}) 
\to\,$U$(N_f^{})\times$U$(N_f)$}.

Other symmetry breaking patterns, particularly involving the possibility of magnetic as well as Cooper-like pairing 
instabilities, have been investigated in Refs.~\cite{Herbut, Khveshchenko2}.


\subsection{Effective action and probability measure}
\label{subsec:Nosignproblem}

The partition function corresponding to Eq.~(\ref{SE}) is given by
\beq
{\mathcal Z}=
\int{\mathcal D}\!A^{}_0{\mathcal D}\psi^{}{\mathcal D}\bar{\psi}^{} \,
\exp(-S^{}_E[\bar{\psi}^{}_a,\psi^{}_a,A^{}_0]),
\eeq
where it is possible to integrate out the fermionic degrees of freedom, as $S_E^{}$ is quadratic in the $\psi_a^{}$. We thus 
obtain
\beq
{\mathcal Z}=
\int{\mathcal D}\!A^{}_0 \,
\exp(-S^g_E[A^{}_0]) \, \det(D[A_0^{}])^{N^{}_f}_{},
\eeq
where 
\beq
S^g_E = \frac{1}{2g^2} \int d^3x\,dt \: (\partial_i^{} A_0^{})^2
\eeq
is the pure gauge part of the action. It is of central importance for the convergence of the Monte Carlo algorithm that the 
above determinant has a definite sign, independently of any particular configuration of the gauge field $A^{}_0$. One way to 
prove that this property is satisfied is to choose a specific representation of the Dirac matrices, such as 
Eq.~(\ref{gammamatrices}), in terms of which $D[A^{}_0]$ can be written as
\beq
D[A^{}_0] = \left(\begin{array}{cc} 
	M[A^{}_0] & 0 \\ 
	0 & -M[A^{}_0] 
	\end{array}\right)
	=
	\left(\begin{array}{cc} 
	M[A^{}_0] & 0 \\ 
	0 & M^\dagger[A^{}_0] 
	\end{array}\right),
\eeq
where 
\beq
M[A^{}_0] = \sigma_0^{} (\partial_0^{} + iA_0^{}) + v\sigma_i^{} \partial_i^{}, \quad i=1,2,
\eeq
and use the facts that $A_0^{}$ is real, and that the Pauli matrices and the momentum operator are Hermitian. The 
latter implies $\partial_\mu^\dagger = -\partial^{}_\mu$, and therefore
\beq
\det(D) = \det(M) \det(M^{\dagger}) = |\det(M)|^2 > 0
\eeq
which, furthermore, is not affected by the introduction of a parity invariant mass term such as Eq.~(\ref{eq:massterm}). 
However, the positivity of $\det(D)$ breaks down in the presence of a chemical potential, which can be thought of as a 
uniform, imaginary contribution to the $A_0^{}$ field.

The fact that $\det(D)$ is positive definite allows for the definition of an effective gauge action that is purely 
real, given by
\beq
\label{Seff}
S_{\text{eff}}[A_0^{}] = -N_f^{}\ln\det(D[A_0^{}]) + S^g_E[A_0^{}],
\eeq
so that the partition function becomes
\beq
\label{Z_eff}
{\mathcal Z} = \int{\mathcal D}\!A^{}_0 \,
\exp(-S^{}_{\text{eff}}[A^{}_0]),
\eeq
where $P[A^{}_0] = \exp(-S^{}_{\text{eff}}[A^{}_0]) > 0$ can be interpreted as a positive definite probability measure 
for a Monte Carlo calculation, as outlined in Section~\ref{sec:lattice}.


\subsection{Operator expectation values
\label{subsec:ExpectationValues}}

The expectation value of a given operator $O[\bar \psi,\psi]$ dependent on the fermion fields can be calculated by taking 
functional derivatives of the generating functional
\bea
{\mathcal Z[\bar\eta, \eta]} &=&
\int{\mathcal D}\!A^{}_0 {\mathcal D}\psi^{}{\mathcal D}\bar{\psi} \nonumber \\
&& \times\:\exp(-S^{}_{E}[A^{}_0,\bar\psi, \psi, \bar\eta, \eta]), \quad
\eea
where source terms have been added to the original action according to
\bea
S_E^{} [A^{}_0,\bar\psi, \psi, \bar\eta, \eta] &=& S_E^{}[A^{}_0,\bar\psi, \psi]
 + \int d^2x dt \, (\bar\psi \eta + \text{H.c.}),  \nonumber \\
\eea
such that the modified effective gauge action is a functional of $A_0^{}$ as well as of 
the sources $\eta, \bar\eta$. It is again possible to integrate out the fermionic degrees of freedom and take 
functional derivatives with respect to the sources in the resulting expression,
\bea
{\mathcal Z[\bar\eta, \eta]} &\propto& 
\int{\mathcal D}\!A^{}_0 \,
\exp(-S^{}_{\text{eff}}[A^{}_0]) \nonumber \\
&& \times\:\exp\left(-\int d^2x dt \,\bar\eta D^{-1}[A_0^{}] \eta\right), \quad\quad
\eea
which makes it possible to obtain expectation values in terms of a path integral over $A_0^{}$ only. While this 
procedure is completely general, it is possible to employ a slightly different approach in order to facilitate the 
computation of the chiral condensate and susceptibility.

The chiral condensate $\sigma$, which is the order parameter of the semimetal-insulator phase transition in graphene, 
is defined by
\beq
\sigma \equiv \langle \bar{\psi}_b^{} \psi_b^{} \rangle,
\eeq
where the fermion fields are evaluated at the same space-time point. It is useful to note that the mass $m_0^{}$ plays the 
role of a source, coupled to $\bar\psi_b^{} \psi_b^{}$. The expectation value of this operator can therefore be obtained 
by first differentiating the partition function with respect to $m_0^{}$ and dividing by the volume, giving 
\bea
\sigma 
&=& \frac{1}{V \mathcal Z}\int{\mathcal D}\!A_0^{}{\mathcal D}\psi^{}{\mathcal D}\bar{\psi}
\int dx \,\bar{\psi}_b^{}(x) {\psi}_b^{}(x) \, \exp(-S_E^{}) \nonumber \\
&=& \frac{1}{V}\frac{\partial \ln {\mathcal Z}}{\partial m_0^{}},
\eea
where we have used the fact that space is homogeneous and therefore the volume average of 
$\bar{\psi}_b^{}(x) {\psi}_b^{}(x)$ can be replaced by its value at an arbitrary point $x$. On the other hand, once the 
fermions have been integrated out, the derivative with respect to $m_0^{}$ yields
\bea
\sigma &=&
\frac{1}{V \mathcal Z}\int{\mathcal D}\!A_0^{} \, \text{Tr}(D^{-1}[A_0^{}]) \, 
\exp(-S_{\text{eff}}[A_0^{}]) \nonumber \\ 
&=& \frac{1}{V}\left\langle \text{Tr}(D^{-1}[A^{}_0]) \right\rangle,
\eea
where the identities
\bea
\det(D[\lambda]) &=& \exp(\text{Tr}(\log(D[\lambda])), \quad\quad \\
\frac{\partial\det(D[\lambda])}{\partial \lambda} &=& 
\det(D[\lambda]) \, \text{Tr}\left(D^{-1}[\lambda] \, \frac{\partial D}{\partial \lambda} \right), \quad\quad
\eea
have been used. The chiral susceptibility $\chi_l^{}$ may be found by taking one more derivative with respect to $m_0^{}$, 
giving
\bea
\chi_l^{} &\equiv& \frac{\partial \sigma}{\partial m_0^{}} \\ \nonumber
&=& 
\frac{1}{V} \left [ 
\left\langle \text{Tr}^2(D^{-1}) \right\rangle - 
\left\langle \text{Tr}(D^{-2}) \right\rangle - 
\left\langle\text{Tr}(D^{-1}) \right\rangle^2 
\right ],
\eea
which is expected to diverge at a second-order phase transition, and may also 
yield constraining information on the universal critical exponents of the transition.


\section{Graphene on the lattice}
\label{sec:lattice}

In this section we formulate the lattice version of Eq.~(\ref{SE}). We begin by discretizing the pure gauge sector, where 
the requirement of gauge invariance implies the use of ``link variables'' to represent the gauge degrees of freedom. The 
``staggered'' discretization of the fermionic sector is then outlined, as it is the preferred choice to 
represent fermions with chiral symmetry at finite lattice spacing.

\subsection{Gauge invariance and link variables}
\label{subsec:gauge}

Recall that the pure gauge part of the Euclidean action is given by
\beq
S^g_E = \frac{1}{2g^2}\int d^3x dt \, (\partial_i^{} A_0^{})^2,
\label{Sgauge}
\eeq
which can be thought of as the non-relativistic limit of the Lorentz-invariant form $\frac{1}{4} 
F_{\mu\nu}^{} F^{\mu\nu}_{}$ where $F_{\mu\nu}^{} = \partial_\mu^{} A_\nu^{} - \partial_\nu^{} A_\mu^{}$, such that
\bea
F_{\mu\nu}^{} F^{\mu\nu}_{} &=& F_{0j}^{} F^{0j}_{} + F_{ij}^{} F^{ij}_{} + F_{i0}^{} F^{i0}_{} \nonumber \\ 
&=& 2 F_{0j}^{} F^{0j}_{} = 2 (\partial_j^{} A_0^{})^2,
\eea
where we have used $F_{ij}^{} = 0$ (no magnetic field) and $\partial_0^{} A_j^{} = 0$ (no electric field induction by a 
magnetic field), valid in the non-relativistic limit ($v \ll c$). Thus, for graphene the only non-vanishing 
contribution is the electric field $E_j^{} = -\partial_j^{} A_0^{}$, which represents the instantaneous Coulomb interaction 
between the quasiparticles.

The action~(\ref{Sgauge}) is invariant under the time-dependent, spatially uniform gauge transformations
\bea
A_0^{} &\rightarrow& A_0^{} + \alpha(t), \nonumber \\  
\psi &\rightarrow& \exp\left\{i \int_0^t dt' \alpha(t') \right\} \psi,
\eea
where $\alpha(t)$ is a function of time only. Thus, in spite of its apparent simplicity, the effective theory of graphene 
possesses a truly local gauge invariance, which should be respected by the lattice action. To this end, one introduces 
temporal link variables
\beq
U_{0,{\bf n}}^{} = U_{\bf n}^{} \equiv \exp\left(i\theta_{\bf n}^{}\right),
\eeq
where $\theta_{\bf n}^{}$ is the dimensionless lattice gauge field evaluated at the lattice point ${\bf n} = 
(n_0,n_1,n_2,n_3)$. The spatial link variables
\beq
U_{i,{\bf n}}^{} = 1,
\eeq
are set to unity. It is convenient to express the discretized version of Eq.~(\ref{Sgauge}) in terms of 
``plaquette'' variables, defined by
\beq
U_{\mu\nu,{\bf n}}^{} = 
U_{\mu,{\bf n}}^{} U_{\nu,{\bf n} + {\bf e}_\mu^{}}^{} 
U^\dagger_{\mu,{\bf n} + {\bf e}_\nu^{}} U^\dagger_{\nu,{\bf n}},
\eeq
where, in the present case of a pure Coulomb interaction, the only non-trivial components are $U_{0i}^{}$ and $U_{i0}^{}$. 
Those plaquette components then correspond to the discretized formulation of the electric field. The remaining components
corresponding to the magnetic field are equal to unity. These statements can be summarized in the expression
\bea
U_{\mu\nu,{\bf n}}^{} 
&=& 
\delta_{\mu 0}^{} \delta_{\nu i}^{} \, U_{\bf n}^{} U^{\dagger}_{{\bf n}+{\bf e}_i^{}}
+ \,\delta_{\nu 0}^{} \delta_{\mu i}^{} \, U^{\dagger}_{\bf n} U_{{\bf n}+{\bf e}_i^{}}^{} \nonumber \\
&&+\: \delta_{\mu 0}^{} \delta_{\nu 0}^{} + \delta_{\mu i}^{} \delta_{\nu j}^{}.
\eea

In terms of the gauge link variables and plaquettes, the discretized gauge action corresponding to Eq.~(\ref{Sgauge})
is given by~\cite{Rothe}
\beq
S^g_E = \beta \sum_{\bf n} \sum_{\mu>\nu}{\left [1 -
\frac{1}{2}\left(U_{\mu\nu,{\bf n}}^{} + U^{\dagger}_{\mu\nu,{\bf n}}\right) \right]},
\label{Sgauge2}
\eeq
where $\beta = 1/g^2$, such that $\beta \to v/g^2$ when the rescaling of Eq.~(\ref{resc}) is applied. In 
Eq.~(\ref{Sgauge2}), the only non-vanishing contributions arise from the terms with 
$(\mu,\nu) = (1,0);(2,0);(3,0);(2,1);(3,1)$ and $(3,2)$. Equation~(\ref{Sgauge2}) may be simplified to
\beq
S^g_{E,C} = \beta \sum_{\bf n} {\left [3 - \sum^3_{i=1} \Re 
\left(U_{\bf n}^{} U^\dagger_{{\bf n}+{\bf e}_i^{}}\right) \right]},
\label{SgaugeC}
\eeq
where $\Re(x)$ denotes the real part of $x$.
Equation~(\ref{SgaugeC}) is referred to as the compact formulation of the discretized gauge action. This formulation is 
known~\cite{compact} to be suboptimal, as compared to the non-compact formulation, for lattice simulations of QED and 
related theories. However, the non-compact formulation may be obtained from Eq.~(\ref{SgaugeC}) by expanding 
$\Re(U_{\bf n}^{} U^\dagger_{{\bf n}+{\bf e}_i^{}})$ to second order in $\theta$,
\beq
\Re\left(U_{\bf n}^{} U^\dagger_{{\bf n}+{\bf e}_i^{}}\right) = 
1 - \frac{1}{2}\left ( \theta_{{\bf n}+{\bf e}_i^{}}^{} - \theta_{\bf n}^{} 
\right)^2 + \:\ldots
\eeq
whereupon the non-compact lattice gauge action is given by
\beq
S^g_{E,N} = \frac{\beta}{2} \sum_{\bf n} {\sum^3_{i=1} \left ( \theta_{{\bf n}+{\bf e}_i^{}}^{} - 
\theta_{\bf n}^{} \right)^2 }.
\label{SgaugeN}
\eeq
Here, and throughout the rest of this paper, we have set the lattice spacing to equal unity, and it is thus dropped from 
all expressions. All dimensionful quantities should therefore be regarded as expressed in units of the 
lattice spacing.


\subsection{Staggered fermions}
\label{subsec:staggeredfermions}

While the discretization of the gauge sector is relatively straightforward, the inclusion of dynamical fermions on the 
lattice is a notoriously difficult problem. One of the main issues when simulating fermions on the lattice is the so-called 
doubling problem (for an overview, see Ref.~\cite{Rothe}, Chapter 4). This problem is related to the chiral invariance of 
the fermionic sector, and arises due to the appearance of multiple (unwanted) zeros in the inverse propagator. In other 
words, one is simulating more fermion flavors than expected, the exact number being dependent on the dimensionality of the 
theory. There exists a number of ways to avoid the doubling problem, but all of them break chiral 
invariance in one way or the other, an inevitable fact encoded in the Nielsen-Ninomiya theorem~\cite{NielsenNinomiya}. The 
solution we have chosen for our simulations of graphene is the ``staggered'' fermion representation 
of Ref.~\cite{Kogut-Susskind}. This choice is optimal for the study of spontaneous chiral symmetry breaking in graphene, as 
it yields the correct number of degrees of freedom while also partially preserving the original chiral symmetry of the 
theory, as will be shown in this section.

In order to discretize the fermionic sector of Eq.~(\ref{SE}) in a way amenable to computer simulations, there are a number 
of choices that need to be made. As a first step, the fermions are integrated out, and the problem is formulated using the 
partition function written purely in terms of the gauge field [Eq.~(\ref{Z_eff})]. The fermions are then represented 
exclusively through the determinant of the Dirac operator $D$. One can attempt to compute the determinant exactly for a 
given $\theta$ configuration, which is feasible due to the low dimensionality of the problem, and is what we have done for 
part of our calculations. Alternatively one may use the so-called pseudofermion method, which we will briefly explain in the 
next section.

In order to arrive at the staggered fermion formulation, a useful starting point is the ``na\"{i}vely'' discretized 
action
\beq
S^f_E[\bar{\psi},\psi,\theta] = 
-\sum_{{\bf n},{\bf m}} 
{\bar \psi}_{b,{\bf n}}^{} \, D_{{\bf n},{\bf m}}^{}[\theta] \,
{\psi}_{b,{\bf m}}^{},
\eeq
where
\bea
D_{{\bf n},{\bf m}}^{}[\theta] &=&
\frac{1}{2} \gamma_0 (\delta_{{\bf n}+{\bf e}_0^{},{\bf m}}^{} \, U_{\bf n}^{} - 
\delta_{{\bf n}-{\bf e}_0^{},{\bf m}}^{} \, U^{\dagger}_{\bf m})\nonumber \\ 
&& +\: \frac{v}{2} \sum_i \gamma_i (\delta_{{\bf n}+{\bf e}_i^{},{\bf m}}^{} - 
\delta_{{\bf n}-{\bf e}_i^{},{\bf m}}^{}) \nonumber \\
&& +\: m_0^{} \delta_{{\bf n},{\bf m}}^{},
\label{D_one}
\eea
with $U_{\bf n}^{} = \exp(i\theta_{\bf n}^{})$. It should be noted that for small $m_0^{}$, Eq.~(\ref{D_one}) becomes 
ill-conditioned, such that the ``chiral limit'' $m_0^{} \to 0$ has to be reached by extrapolation. The boundary conditions 
of the fermion fields are periodic in 
the spatial directions and anti-periodic in the temporal direction. It is possible, using a local unitary transformation on 
the fermion fields, to simultaneously diagonalize the Dirac matrices in Eq.~(\ref{D_one}) and thereby decouple the spinor 
components. This procedure, known as the Kawamoto-Smit transformation~\cite{KawamotoSmit} or simply as 
``spin-diagonalization'', is defined by
\bea
\psi_{\bf n}^{} &\rightarrow& T_{\bf n}^{} \, \chi_{\bf n}^{}, \nonumber \\
\bar\psi_{\bf n}^{} &\rightarrow& \bar\chi_{\bf n}^{} \, T^\dagger_{\bf n},
\eea
which in the Dirac operator~(\ref{D_one}) effects the transformation
\beq
\gamma^\mu \to T^\dagger_{\bf n} \, \gamma^\mu \, T_{{\bf n}+{\bf e}_\mu^{}}^{}
\eeq
on the Dirac matrices $\gamma^\mu$. The transformed fermion fields $\chi_{\bf n}^{}$ are referred to
as staggered spinors. It is straightforward to show that 
the choice $T_{\bf n}^{} = \gamma^{n_0^{}}_0 \gamma^{n_1}_1 \gamma^{n_2}_2$ satisfies
\bea
T^\dagger_{\bf n} \, \gamma^\mu \, T_{{\bf n}+{\bf e}_\mu^{}}^{} \!\!\! &=& \eta^\mu_{\bf n} \, \openone,
\eea
where the Kawamoto-Smit phases are given by
\bea
\eta^0_{\bf n} &=& 1, \nonumber \\
\eta^1_{\bf n} &=& (-1)^{n_0^{}}_{}, \nonumber \\
\eta^2_{\bf n} &=& (-1)^{n_0^{} + n_1^{}}_{}.
\eea
In this fashion the Dirac structure is removed, resulting in a sum of four identical terms in the action, one for each 
component of the original four-component Dirac spinor $\psi_{\bf n}^{}$. These copies are referred to as staggered 
flavors. It has been shown in Ref.~\cite{BurdenBurkitt} that for each staggered flavor one recovers, in the continuum limit, 
two four-component Dirac flavors. Thus, by retaining one staggered flavor, it is possible to have exactly eight continuum 
fermionic degrees of freedom, which is the correct number for graphene. The action of a single staggered flavor
is given by
\beq
S^f_E[\bar{\chi},\chi,\theta] = 
-\sum_{{\bf n},{\bf m}} 
{\bar\chi}_{\bf n}^{} \, K_{{\bf n},{\bf m}}^{}[\theta] \,
{\chi}_{\bf m}^{},
\eeq
where the staggered Dirac operator is
\bea
K_{{\bf n},{\bf m}}^{}[\theta] &=&
\frac{1}{2}(\delta_{{\bf n}+{\bf e}_0^{},{\bf m}} \, U_{\bf n}^{} - 
\delta_{{\bf n} - {\bf e}_0^{},{\bf m}}^{} \, U^{\dagger}_{\bf m}) \nonumber \\
&& +\: \frac{v}{2} \sum_i \eta^i_{\bf n} (\delta_{{\bf n}+{\bf e}_i^{},{\bf m}}^{} - 
\delta_{{\bf n}-{\bf e}_i^{},{\bf m}}^{}) \nonumber \\
&& +\: m_0^{} \delta_{{\bf n},{\bf m}}^{}.
\label{Kstag}
\eea
The operator $K$ thus replaces $D$ in all expressions for the probability, chiral condensate and susceptibility that were 
derived in the previous sections. As expected from the Nielsen-Ninomiya theorem, the staggered lattice action does not 
retain the full U$(4)$ chiral symmetry of the original graphene action at finite lattice spacing. As shown in 
Ref.~\cite{BurdenBurkitt}, only a subgroup U$(1)\times$U$(1)$ remains upon discretization. Spontaneous condensation of 
$\bar\chi \chi$, or equivalently the introduction of a parity invariant mass term, reduces this symmetry to U$(1)$. The 
focus of this work is on the phase transition associated with such a chiral symmetry-breaking pattern.

Finally, it should be pointed out that the situation 
concerning graphene is unusually favorable, in the sense that the staggered formalism somewhat fortuitously provides the 
correct number of fermionic degrees of freedom as $N_f^{} = 2$ for graphene monolayers. In general, staggered fermions 
provide only a compromise solution in the sense that some degree of chiral symmetry is preserved, at the price of retaining 
some of the doubling originally present in the discretized fermion action. Indeed, if the case of $N_f^{} = 1$ were to be 
simulated, it would be necessary to resort to the uncontrolled and controversial ``rooting'' trick~\cite{rooting}, whereby 
the desired number of continuum flavors is restored by taking the appropriate root of the Dirac operator.


\subsection{Computation of observables}

The computation of $\sigma$ and $\chi_l^{}$ from ensembles of gauge field configurations necessitates, in principle, the 
full inversion of $K$ and $K^2$. Such a procedure may potentially become extremely time-consuming for large lattices. In 
this respect, a choice exists between direct sparse solvers, such as PARDISO~\cite{PARDISO}, and iterative solvers of the 
conjugate gradient type, such as BiCGStab~\cite{BiCGStab}. As the lattice size is increased, the performance of the direct 
solver scales much worse than the iterative solver, by a factor roughly proportional to the lattice volume. Nevertheless, 
the direct sparse solvers remain an attractive choice for a number of reasons: the performance of a direct solver is 
independent of the condition number of $K$, which is the ratio of its largest and smallest eigenvalues, and this is 
particularly significant close to a transition and for small $m_0^{}$. Furthermore, direct solvers feature optimized 
parallelization and are efficient at handling inversion problems with multiple right-hand-sides. In view of this, PARDISO 
has been found to be the solver of choice for the efficient computation of observables on the presently used lattice 
volumes.

Regardless of the type of solver used, the full inversion of $K$ quickly becomes impractically expensive when the lattice 
size is increased. In this situation, it is possible to resort to a stochastic estimator~\cite{Bitaretal}, which constitutes 
an alternative to the exact calculation of $\text{Tr}(K^{-1})$. A suitable stochastic estimator for $\sigma$ is given by
\beq
\hat\sigma = \frac{1}{V}\sum_{{\bf n}, {\bf m}} \xi^\dagger_{\bf n} \,
K^{-1}_{{\bf n},{\bf m}} [\theta] \, \xi_{\bf m}^{},
\label{eq:StochasticEstimator}
\eeq
where the $\xi_{\bf n}^{}$ are random Gaussian variables which satisfy $\langle\langle \xi_{\bf n}^{} \rangle\rangle = 0$ 
and $\langle\langle \xi_{\bf m}^\dagger \xi_{\bf n}^{} \rangle \rangle = \delta_{{\bf m},{\bf n}}^{}$, where the double 
bracket notation indicates an average over $\xi_{\bf n}^{}$. 

For a given gauge configuration, averaging Eq.~(\ref{eq:StochasticEstimator}) over $\xi_{\bf n}^{}$ yields 
$\text{Tr}(K^{-1})$, which only requires application of the inverse to a limited number of random Gaussian vectors. With 
this approach it is also straightforward to compute $\text{Tr}(K^{-2})$, by simply applying the inverse to each random 
vector one more time. Adequate accuracy for $\sigma$ and $\chi_l^{}$ is achieved using $\sim 100$ random vectors for each 
gauge configuration, independently of the lattice volume used.


\section{Monte Carlo Strategies}
\label{sec:MC}

This section presents the two Monte Carlo algorithms that we have used to study the discretized low-energy effective theory 
of graphene. We begin by outlining the Metropolis Monte Carlo algorithm which, although conceptually simpler, becomes 
computationally inefficient beyond a certain lattice volume, after which we proceed to describe the more advanced and highly 
efficient approach involving the Hybrid Monte Carlo~(HMC) algorithm with pseudofermions.


\subsection{Metropolis Monte Carlo}
\label{subsec:MMC}

As shown in Section~\ref{subsec:Nosignproblem}, the structure of the fermion determinant allows for a positive 
definite probability measure. Indeed, as shown in Section~\ref{subsec:ExpectationValues}, an effective action can be defined 
such that expectation values of observables can be written as averages over field configurations weighted by
\beq
P[\theta] \equiv \exp(-S_\text{eff}^{}[\theta]) = \det(K[\theta]) \, \exp(-S^g_E[\theta]),
\eeq
where the matrix $K$ corresponds to the staggered Dirac operator of Eq.~(\ref{Kstag}).
In the Metropolis algorithm~\cite{Metropolis}, a given gauge field configuration $\theta$ is updated by the introduction of 
a small change at a randomly chosen lattice site. The updated configuration $\theta'$ is then accepted with probability
\bea
\label{eq:MetropolisTest}
p &\equiv& \frac{P[\theta']}{P[\theta]} = \exp(-\Delta S), \nonumber \\
\Delta S &=& S_{\text{eff}}^{}[\theta'] - S_{\text{eff}}^{}[\theta].
\eea
If the new configuration $\theta'$ is rejected, $\theta$ is retained, and a new change proposed. In this fashion, a 
so-called Markov chain of gauge configurations is generated, in which the samples are distributed according to the desired 
probability measure. After an appropriate number of thermalization steps, gauge configurations can be saved at regular 
intervals, which should allow for adequate decorrelation. The central limit theorem then 
guarantees that for $\mathcal N$ uncorrelated samples, the statistical uncertainties will decrease as $1/\sqrt{\mathcal N}$. 
The decorrelation can be measured in terms of the number of full sweeps of the lattice required between two consecutive 
observations, in order for the autocorrelation of the ensemble of gauge configurations to become insignificant. For the 
Metropolis algorithm, a proper balance between update size and decorrelation is achieved for acceptance rates of $\sim 
60-70\%$.

In spite of its simplicity, the Metropolis approach has several inherent disadvantages. The most serious one arises as the 
fermion action is non-local, in the sense that updating a single lattice site requires a full recalculation of 
$\det(K)$. This disadvantage is exacerbated by the fact that decorrelation is dependent on the number of full sweeps of the 
lattice, and the number of sites to be updated increases as the lattice size is increased. Even with highly efficient 
parallel sparse solvers, the execution time scales as $\sim V^3$, such that it is bound to become impractical above a certain 
maximum lattice size. Also, as the updates in the Metropolis algorithm are entirely random, it is usually only possible to 
update very few lattice sites at once without ruining the acceptance rate. In the Sec.\ref{subsec:HMC}, we give an overview of 
the HMC algorithm, which is designed to overcome these difficulties.


\subsection{Hybrid Monte Carlo}
\label{subsec:HMC}

The problem of efficient updating of the gauge field in theories with dynamical fermions has been addressed in 
Ref.~\cite{DKPR} where the Hybrid Monte Carlo~(HMC) algorithm was introduced. In this approach, the gauge field is evolved 
deterministically along a Molecular Dynamics~(MD) trajectory, such that the entire lattice is updated at once. Thus, the 
number of updates required for decorrelation within the HMC algorithm is dramatically reduced, although the number of MD 
trajectories required for decorrelation roughly equals the number of sweeps necessary in the Metropolis approach.

The basic idea of the HMC algorithm is to evolve a given initial configuration $\theta_{\bf n}^{}$ in a fictitious time 
$\tau$ according to the classical equations of motion, with a Hamiltonian given by
\beq
H = \sum_{\bf n}\frac{\pi_{\bf n}^2}{2} + S_E^{}[\theta]
\label{HMC_ham}
\eeq
where $S_E[\theta]$ is the Euclidean action to be sampled, and $\pi_{\bf n}^{}$ is a momentum conjugate to 
$\theta_{\bf n}^{}$. This momentum is introduced as an auxiliary field, with the sole purpose of defining the above 
dynamics. The field $\pi_{\bf n}^{}$ is of no consequence to the path integral that defines the theory, as its contribution 
factors out completely. It has been shown in Ref.~\cite{DKPR} that the procedure of classically evolving 
$(\theta_{\bf n}^{},\pi_{\bf n}^{}) \to (\theta'_{\bf n},\pi'_{\bf n})$ using the above Hamiltonian, and choosing the 
initial $\pi_{\bf n}^{}$ from a random Gaussian distribution, produces a Markov chain of gauge field configurations
distributed according to the desired probability measure. 

Because the MD evolution is in principle exact, a trajectory that is long enough should provide the desired decorrelation 
between consecutive samples, provided that the pseudofermion field is refreshed at regular intervals. Ideally, a $100\%$ 
acceptance rate should thus be achievable. In practice, however, the MD evolution is implemented with a finite time step 
$\Delta\tau$, which introduces a systematic error. However, as long as the evolution remains reversible, the effects of that 
error on the distribution of gauge field configurations can be eliminated by means of a Metropolis step, comparing the 
initial and final configurations after each MD evolution, where Eq.~(\ref{HMC_ham}) plays the r\^ole of the effective action 
in Eq.~(\ref{eq:MetropolisTest}).

While the HMC algorithm achieves very efficient updating of the gauge field, a potentially serious drawback is that the 
updating procedure requires (in principle) the full evaluation of $K^{-1}$ which is computationally prohibitively 
expensive, even more so than $\det(K)$. Because of this, a number of methods have been developed that seek to circumvent the 
necessity of calculating $K^{-1}$. In one of these, the so-called R-algorithm~\cite{Gottlieb}, the inverse is approximated 
by a stochastical estimator which, however, introduces a systematical error due to the loss of reversibility. Arguably, the 
method of choice is the $\Phi$-algorithm~\cite{Gottlieb}, which reduces the MD evolution into a sparse operation by 
re-expressing the square of the fermion determinant as a path integral over complex scalar fields known 
as pseudofermions, while simultaneously maintaining the desirable features of the HMC approach.


\subsection{Pseudofermions}
\label{subsec:pseudofermions}

As the pseudofermion method is explained in great detail elsewhere (for pedagogical reviews, see 
Refs.~\cite{Rothe,DeGrandDeTar}) we shall only concern ourselves with outlining the basic idea, which is based on the 
identity
\beq
\det(Q) \propto
\int {\mathcal D} \phi^\dagger {\mathcal D} \phi
\, \exp(-S_E^p),
\eeq 
where the constant of proportionality is of no consequence. Here, $\phi, \phi^\dagger_{}$ are pseudofermion fields (which 
are bosonic but nevertheless satisfy anti-periodic boundary conditions in the temporal direction), $Q \equiv K^\dagger 
K$ and the pseudofermion action is
\beq
S_E^p = \sum_{{\bf n},{\bf m}} \phi^\dagger_{\bf n} \, Q^{-1}_{{\bf n},{\bf m}}[\theta] \, \phi_{\bf m}^{}
= \sum_{\bf n} \xi^\dagger_{\bf n} \xi_{\bf n}^{},
\eeq
where $\xi$ follows a Gaussian distribution, related to the pseudofermion field by $\phi = K^\dagger \xi$. 

In order to simulate graphene, one requires $\det(K)$, not $\det(Q) = \det(K^\dagger K)$. Thus, using the pseudofermions 
according to the above prescription effectively doubles the number of degrees of freedom. Fortunately, the staggered fermion 
action allows for an odd-even decomposition~\cite{DeGrandDeTar}, such that a single staggered flavor can be simulated. In 
the odd-even decomposition, the lattice is separated into sublattices of even and odd sites, according to the sign of 
$(-1)^{n_0^{} + n_1^{} + n_2^{}}_{}$. Thus, as the derivative operator connects odd (even) sites with even (odd) ones, while 
the mass term connects odd (even) sites with odd (even) ones, the following odd-even decomposed form results:
\beq
K = \left ( \begin{array}{cc}
	 m_0^{} & K_{oe}^{} \\ 
	 K_{eo}^{} & m_0^{} 
	\end{array}
	 \right ),
\eeq
and therefore 
\beq
Q = \left ( \begin{array}{cc}
	 K^\dagger_{eo}K_{oe}^{} + m^2_0 & 0 \\ 
	 0 & K^\dagger_{oe}K_{eo}^{} + m^2_0
	\end{array}
	 \right ),
\eeq
which, using the fact that $K^\dagger_{oe} = - K_{eo}^{}$, has been factorized into blocks of even-even and odd-odd 
elements. As a consequence,
\beq
\det(Q) = \det\left(K^\dagger_{eo}K_{oe}^{} + m^2_0\right)^2_{}.
\eeq
Thus, in order to recover $\det(K)$, it suffices to retain only the even-even (or odd-odd) block of Q. In practice, this is 
implemented simply by discarding either the odd (or even) elements of $\phi$.

In the presence of pseudofermions, the MD Hamiltonian becomes 
\beq
H = \sum_{\bf n}\frac{\pi_{\bf n}^2}{2} + S^{g}_E + S^p_E,
\eeq
and the equations of motion are
\bea
\dot\theta_{\bf n}^{} &=& \frac{\delta H}{\delta \pi_{\bf n}^{}} = \pi_{\bf n}^{}, \\
\dot\pi_{\bf n}^{} &=&  - \frac{\delta H}{\delta \theta_{\bf n}^{}} \equiv F^g_{\bf n} + F^p_{\bf n},
\eea
where the ``force term'' associated with the gauge action takes the form
\bea
F^g_{\bf n} &\equiv& - \frac{\delta S^{g}_E}{\delta \theta_{\bf n}^{} } \\ \nonumber
&=&  -\frac{1}{g^2} \sum_{j=1}^3 \Im\left(U_{\bf n}^{} U^\dagger_{{\bf n}+ {\bf e}_j^{}} 
- U_{{\bf n}-{\bf e}_j^{}}^{} U^\dagger_{\bf n}\right) \\ \nonumber
&=& -\frac{1}{g^2} \left [ 6\,\theta_{\bf n}  
- \sum_{j=1}^3 \left(\theta_{{\bf n}+ {\bf e}_j^{}}  + \theta_{{\bf n}- {\bf e}_j^{}}  \right)  \right] \:+\: ...\, ,
\eea
where $\Im(x)$ is the imaginary part of $x$; the second line in this equation corresponds to the compact formulation and the 
last line, obtained by expanding in powers of $\theta$, shows the result for the non-compact case. The pseudofermion 
contribution is given by
\bea
F^p_{\bf n} &=& - \frac{\delta S^p_E}{\delta \theta_{\bf n}^{} } 
\label{PF_force} \\ 
&=& - \sum_{\bf n} \phi^\dagger \frac{\delta Q^{-1}}{\delta \theta_{\bf n}^{}} \, \phi 
= \sum_{\bf n} \phi^\dagger Q^{-1} \frac{\delta Q}{\delta \theta_{\bf n}^{}} \, Q^{-1} \phi.
\nonumber
\eea
The essence of the $\Phi$-algorithm is the treatment of $\phi$ as a constant background field throughout each MD trajectory. 
After each MD evolution, the pseudofermion field is refreshed using random Gaussian noise according to $\phi = K^\dagger 
\xi$. Computationally, the great advantage of this algorithm is that in each step $\Delta\tau$, the calculational effort is 
reduced to applying the inverse of $K^\dagger K$ to a single vector $\phi$, which is significantly less expensive than 
computing the full inverse.

The numerical integration of the MD equations of motion requires a reversible method, and the usual choice is the leap-frog 
integration formula~\cite{DKPR} which is also area-preserving. The calculation of the pseudofermion force in 
Eq.~(\ref{PF_force}) is preferentially accomplished using an iterative solver such as BiCGStab~\cite{BiCGStab}, in which 
case the algorithm scales roughly as $\sim V$. Nevertheless, in practical calculations the scaling is inevitably somewhat 
worse, as the truncation error of the leap-frog method tends to increase with increasing lattice size, necessitating a 
smaller timestep $\Delta\tau$.

In the present study of the low-energy effective theory of graphene, we have used both the Metropolis and HMC 
algorithms, verifying that for any given set of parameters the results agree within statistical uncertainties. 
We now turn to a presentation of our simulation results. 


\section{Results}
\label{sec:results}

In our simulations, the fermions live in a volume of extent $V \equiv L_x^2 \times L_t^{}$, while the gauge bosons also 
propagate in the $z$-direction of length $L_z^{}$. Increasing $L_z^{}$ beyond~$8$ was found to have no discernible effects. 
The results will thus be referred to by the short-hand notation $L_x^2 \times L_t^{}$. Also, the action~(\ref{SE}) has been 
rescaled according to Eq.~(\ref{resc}), such that $\beta \equiv v/g^2$ and $v = 1$ in the staggered Dirac operator of 
Eq.~(\ref{Kstag}). Our simulations have been performed at finite (but small) values of $m_0^{}$, such that the limit $m_0^{} 
\to 0$ is reached by extrapolation.

We have performed simulations on lattice sizes up to $20^2\times20$ using the Metropolis method and $28^2\times28$ using 
HMC. The former method scales roughly as $V^3$ and therefore quickly becomes uneconomical when the lattice volume is 
increased. However, an advantage of the Metropolis method is that the speed of the algorithm is independent of the condition 
number of the staggered Dirac operator $K$, as the fermionic determinant is evaluated using a direct solver. In contrast, 
the HMC algorithm with pseudofermions scales roughly as $\sim V$, if used together with an iterative solver such as 
BiCGStab~\cite{BiCGStab}. However, the HMC algorithm then becomes sensitive to the condition number of $K$, such that 
obtaining data becomes more difficult at small bare fermion masses or close to the critical coupling. This problem can be 
somewhat alleviated using a direct solver such as \mbox{PARDISO}~\cite{PARDISO}, but in that case the HMC algorithm scales 
roughly as $\sim V^2$.

Within the Metropolis approach, $\sim 240$~uncorrelated configurations were generated for each value of ($\beta,m_0^{}$). 
When using the HMC algorithm, a similar number of MD trajectories were generated for each datapoint. The optimal MD 
time step $\Delta\tau$ was found to be dependent on the values of $\beta$ and $m_0^{}$. In order to simultaneously optimize 
the acceptance rate, decorrelation and execution time, $\Delta\tau$ was adjusted in the range $[0.01,0.03]$, while the 
number of steps $N_\tau^{}$ was chosen randomly from a Poisson distribution such that the average MD trajectory length 
between updates of the pseudofermion field was $\bar\tau = N_\tau^{} \Delta\tau = 2$. The choice of $\bar\tau \simeq 2.5$ 
was found to give optimal decorrelation.

The HMC algorithm is the method of choice for lattices larger than $20^2 \times 20$. As a check on the HMC code, the 
datapoints for $16^2 \times 16$ computed using the Metropolis algorithm in Ref.~\cite{DrutLahde1} were recomputed using the 
HMC method, and found to agree within statistical uncertainties. In all cases, the uncertainties were estimated using the 
Jackknife method~\cite{jackknife}.


\subsection{The semimetal-insulator transition}

\begin{figure}[t]
\includegraphics[width=\columnwidth]{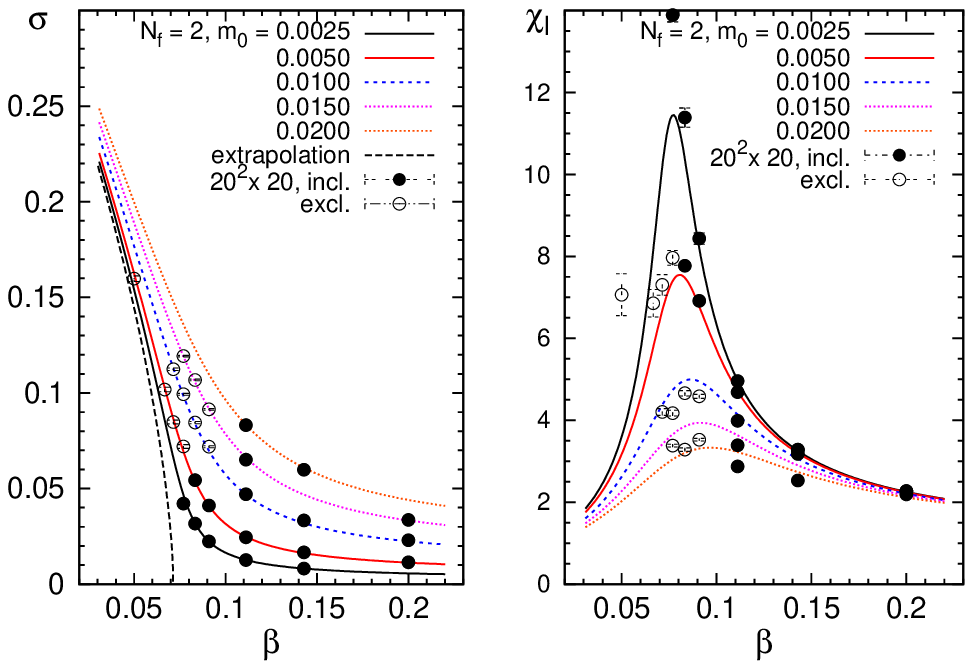}
\includegraphics[width=\columnwidth]{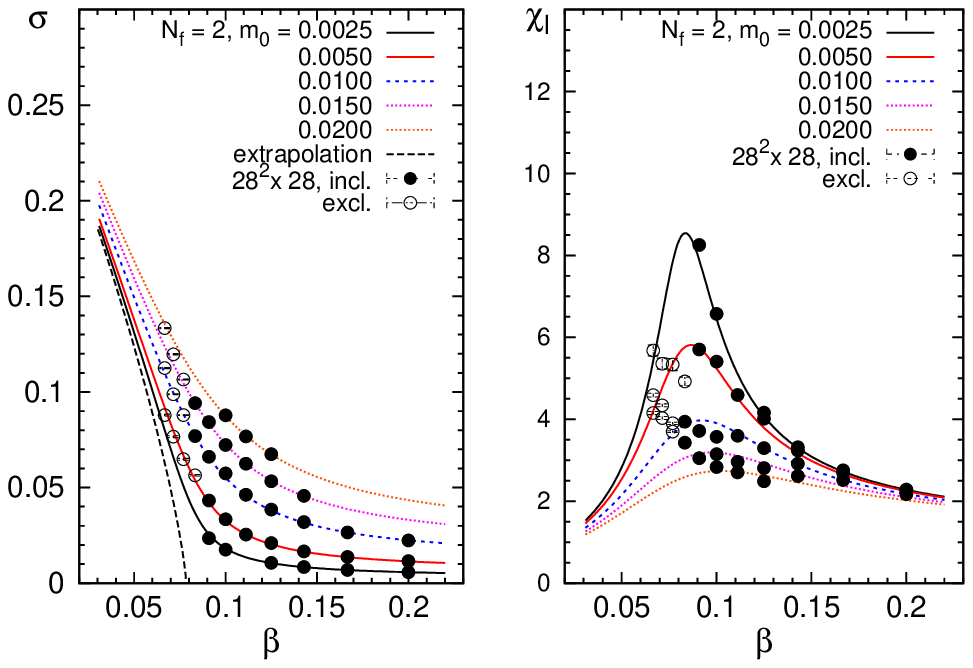}
\caption{\label{fig:chiralcondensate1} (Color online) Chiral condensate (upper left panel) and susceptibility (upper right 
panel) for lattice size $20^2\times20$. Lower panels show the same quantities for $28^2\times28$. The lines represent 
$\chi^2_{}$ fits of Eq.~(\ref{EOS}) to $\sigma$ only, with $X_0^{},X_1^{},Y_1^{}, \delta$ and $\beta_c^{}$ as free 
parameters; the datapoints with largest finite-size effects have been excluded from the fit. The optimal parameter values are: 
for $20^2\times20$, $X_0^{} = 0.665 \pm 0.2$, $X_1^{} = -0.280 \pm 0.088$ and $Y_1^{} = -0.2869 \pm 0.090$, $\delta = 2.27 
\pm 0.13$, $\beta_c^{} = 0.0721 \pm 0.0006$; for $28^2\times28$, $X_0^{} = 0.3427 \pm 0.028$, $X_1^{} = -0.190 \pm 0.014$ 
and $Y_1^{} = -0.179 \pm 0.014$, $\delta = 2.309 \pm 0.037$, $\beta_c^{} = 0.0785 \pm 0.0003$. The uncertainties are purely 
statistical.}
\end{figure}

In order to determine the critical coupling $\beta_c$ for spontaneous chiral symmetry breaking, we calculated the chiral 
condensate $\sigma$ and susceptibility $\chi_l^{}$ for $\beta$ between 0.05 and 0.5, and for $m_0^{}$ between 0.0025 and 
0.020 (in lattice units). Fig.~\ref{fig:chiralcondensate1} shows our data for lattice sizes $20^2\times20$ (upper panels) 
and $28^2\times28$ (lower panels).

The chiral condensate increases as $\beta$ is decreased, more sharply so below $\beta \simeq 0.1$. This behavior becomes 
more pronounced as $m_0$ is decreased, providing the first indication of a phase transition as the Coulomb coupling is 
increased. In turn, the susceptibility also grows sharply around $\beta \simeq 0.1$. This feature tends to disappear for 
$m_0^{} > 0.010$ as the lattice volume is increased. Thus, in order to understand the properties of the transition, masses 
smaller than $m_0 \sim 0.010$ should be used in the simulation. This situation is similar to that encountered in quenched 
QED$_4^{}$~\cite{Kogutetal} where it was concluded that for the critical region to be reached, bare masses smaller than 
$\sim 0.025$ should be used. On the other hand, for the smallest mass of $m_0 = 0.0025$, the change in the susceptibility as 
a function of the lattice volume appears to be relatively mild for $\beta > 0.09$. The rise in the susceptibility is therefore 
likely to be a real feature, indicating that the critical region has been reached.

\begin{figure}[t]
\includegraphics[width=\columnwidth]{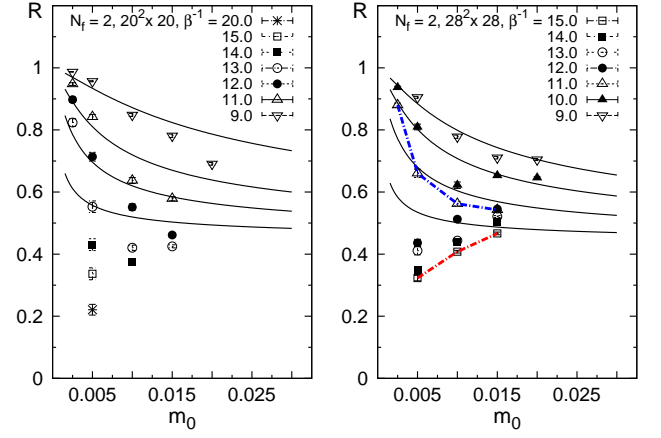}
\caption{\label{fig:R} (Color online) Logarithmic derivative $R$ for lattice sizes $20^2\times20$ (left panel) and 
$28^2\times28$ (right panel). The solid lines for the largest four values of $\beta$ correspond to the restricted fits shown 
in Fig.~\ref{fig:chiralcondensate1}. The dashed red line in the right panel connects the datapoints for 
$\beta = 1/15.0 \approx 0.067$, 
where the downward slope is characteristic of the spontaneously broken phase. On the other hand, the dashed blue line 
connecting the datapoints for $\beta = 1/11.0 \approx 0.091$ clearly indicates that chiral symmetry remains unbroken in the 
limit $m_0^{} \to 0$ for that value of $\beta$. The evidence for spontaneous chiral symmetry breaking is 
significantly stronger for $28^2\times28$, where the data for $\beta = 1/13.0 \approx 0.077$ are consistent with the broken 
phase, while for $20^2\times20$ the opposite is true. The $28^2\times28$ lattice favors a slightly larger value of 
$\beta_c^{}$, while simultaneously disfavoring the classical critical exponent $\delta = 3$.}
\end{figure}

In spite of the compelling qualitative evidence presented above, the nature of the simulational study precludes the use of 
bare masses $m_0^{}$ that are small enough so that the distortion introduced is negligible. What is needed is a controlled 
way of obtaining information about the massless limit, using the data at hand, taken at small but finite $m_0^{}$. A 
suitable observable is provided by the logarithmic derivative $R$~\cite{Kocic:1992pf} of the chiral condensate with respect 
to $m_0^{}$,
\begin{eqnarray}
R &\equiv & \left.\frac{\partial \ln\sigma}
{\partial \ln m_0^{}}\right|_\beta^{}
= \left.\frac{m_0^{}}\sigma \left(
\frac{\partial \sigma}{\partial m_0^{}}\right)\right|_\beta^{},
\label{Rrat}
\end{eqnarray}
which allows for a more precise determination of the critical coupling $\beta_c^{}$, as well as for an estimate of the 
universal critical exponent $\delta$ (see Eq.~\ref{exponents_def}). 
In the limit $m_0^{} \to 0$, $R \to 1$ in the chirally symmetric phase since $\sigma 
\propto m_0^{}$; while at the critical coupling $\beta=\beta_c^{}$, one expects $R \to 1/\delta$. Finally, $R$ vanishes in 
the limit $m_0^{} \to 0$ in the spontaneously broken phase, where $\sigma \neq 0$ for $m_0^{} \to 0$. The data on $R$ in 
Fig.~\ref{fig:R} (right panel) indicate that chiral symmetry is spontaneously broken for 
$\beta = 1/14.0 \approx 0.071$, but remains unbroken for $\beta = 1/11.0 \approx 0.091$. We thus conclude, using the 
$28^2\times 28$ data, that 
\beq
0.071 < \beta_c^{} < 0.091,
\eeq
which could be further refined by use of larger lattice volumes and smaller values of $m_0^{}$. 
%


\subsection{Determination of the equation of state}

While the logarithmic derivative $R$ may provide model-independent information on the critical coupling as well as the 
exponent $\delta$, it involves the chiral susceptibility and is therefore prone to large finite-size effects. A more 
accurate determination of $\beta_c^{}$ can be achieved by means of an appropriate equation of state~(EOS) 
\begin{eqnarray}
m_0^{} = f(\sigma,\beta),
\label{EOSgeneral}
\end{eqnarray}
which is to be fitted to the simulation data on the chiral condensate. This EOS can then yield direct information on 
$\beta_c^{}$ as well as the critical exponents $\delta$ and $\bar\beta$, defined by
\bea
\label{exponents_def}
\delta &\equiv& \left. \left [ \frac{\partial \ln \sigma}{\partial \ln m_0^{}} \right]^{-1}
\right|_{\beta = \beta_c^{}, m_0^{} \to 0}, \\
\bar \beta &\equiv& \left. \frac{\partial \ln \sigma}{\partial \ln (\beta_c - \beta)} 
\right|_{m_0^{} = 0, \beta \nearrow \beta_c^{}}.
\eea
In addition, using the scaling relation
\beq
\bar\beta (\delta - 1) = \gamma
\label{scaling}
\eeq
one can obtain the critical exponent $\gamma$, defined by
\beq
\gamma \equiv -\left. \frac{\partial \ln \chi}{\partial \ln (\beta_c^{} - \beta)} 
\right|_{m_0^{} = 0, \beta \to \beta_c^{}}
\label{gamma}
\eeq
The EOS also provides a means for an extrapolation $m_0^{} \to 0$, which necessitates an {\it ansatz} for 
Eq.~(\ref{EOSgeneral}). We have considered an EOS similar to those successfully applied~\cite{Gockeleretal1,Gockeleretal2,Gockeleretal3,AliKhan:1995wq} 
to QED$_4^{}$,
\begin{eqnarray}
m_0^{}X(\beta) &=& Y(\beta) f_1^{}(\sigma) + f_3^{}(\sigma),
\label{EOS}
\end{eqnarray}
where the functions $X$ and $Y$ are expanded around $\beta_c^{}$ such that
$X(\beta) = X_0^{} + X_1^{}(1-\beta/\beta_c^{})$ and $Y(\beta) = Y_1^{}(1-\beta/\beta_c^{})$.
The dependence of Eq.~(\ref{EOS}) on $\sigma$ is
\begin{eqnarray}
f_1^{}(\sigma) = \sigma^b_{}, \quad  
f_3^{}(\sigma) = \sigma^\delta_{}, 
\label{ffuncs}
\end{eqnarray}
where $b \equiv \delta - 1/\bar\beta$. Thus Eq.~(\ref{EOS}) can be used to study deviations from the classical exponents 
$\delta = 3$ and $\bar \beta = 1/2$. It should be noted that for the case of QED$_4^{}$~\cite{Gockeleretal1,Gockeleretal2,Gockeleretal3,AliKhan:1995wq}, 
an extended version of the {\it ansatz}~(\ref{ffuncs}) has been used to include logarithmic corrections to the EOS.

\begin{figure}[t]
\includegraphics[width=\columnwidth]{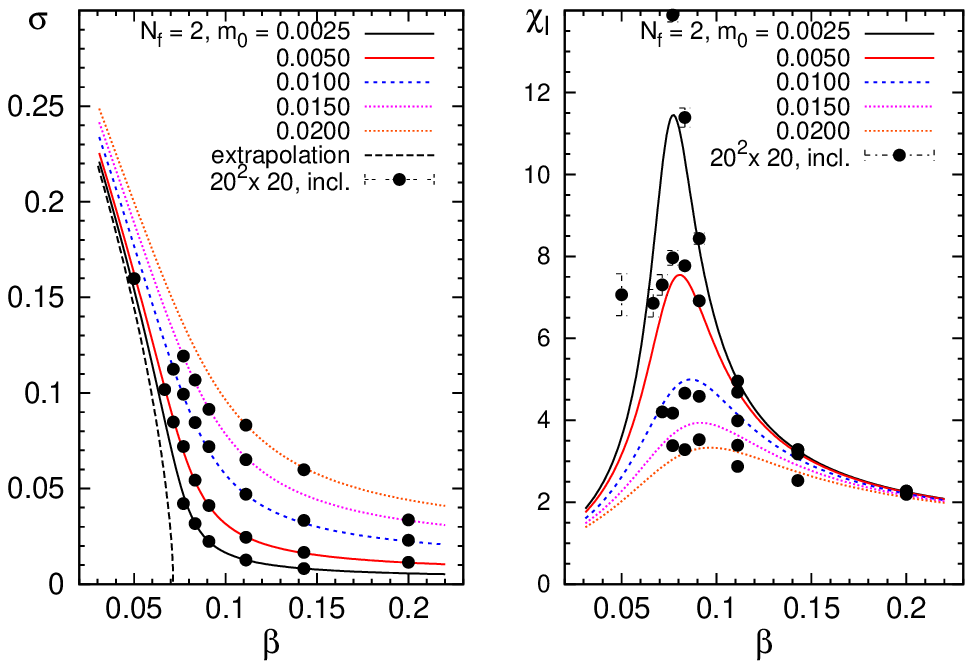}
\includegraphics[width=\columnwidth]{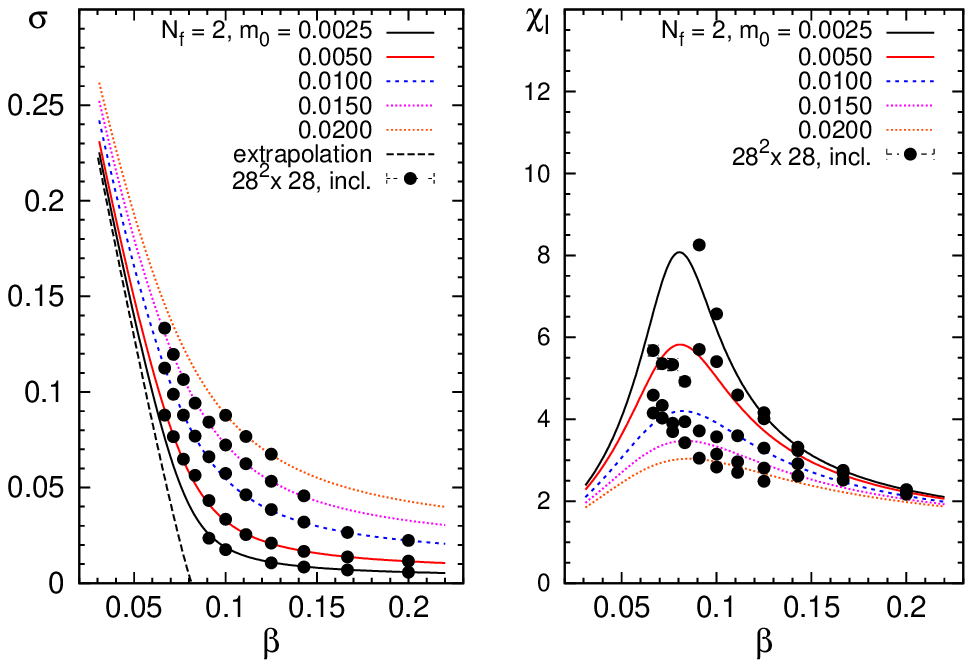}
\caption{\label{fig:chiralcondensate2} (Color online) Chiral condensate (upper left panel) and susceptibility (upper right 
panel) for lattice size $20\times20$. Lower panels show the same quantities for $28\times28$. The lines represent 
$\chi^2_{}$ fits of Eq.~(\ref{EOS}) to $\sigma$ only, with $X_0^{},X_1^{},Y_1^{}, \delta$ and $\beta_c^{}$ as free 
parameters; all datapoints have been included in the fit, regardless of the estimated magnitude of finite-size effects. The 
optimal parameter values are: for $20^2\times20$, $X_0^{} = 0.364 \pm 0.029$, $X_1^{} = -0.156 \pm 0.013 $ and $Y_1^{} = 
-0.159 \pm 0.013$, $\delta = 2.573 \pm 0.041$, $\beta_c^{} = 0.0715 \pm 0.0003$; for $28^2\times28$, $X_0^{} = 0.834 \pm 
0.024$, $X_1^{} = -0.409 \pm 0.011 $ and $Y_1^{} = -0.418 \pm 0.012 $, $\delta = 1.889 \pm 0.017 $, $\beta_c^{} = 0.0815 \pm 
0.0004 $.} 
\end{figure} 

While it is possible to fit both $\sigma$ and $\chi_l^{}$ simultaneously, it is advantageous to use the 
latter quantity as a consistency check only, as the finite-size effects are much smaller for $\sigma$. It is also useful to 
restrict the fit range to the datapoints where such effects are not too large. The results of the fits with restricted range 
are given in Fig.~\ref{fig:chiralcondensate1}, whereas the results of a full fit to all datapoints is shown in 
Fig.~\ref{fig:chiralcondensate2}. The results for $\beta_c^{}$ and $\delta$ are much more consistent for the restricted 
dataset. The fit results for the restricted $28^2\times28$ dataset indicate a critical coupling of $\beta_c^{} = 0.0785 \pm 
0.0003$ and a critical exponent $\delta = 2.309 \pm 0.037$. All of the fits described above have been performed using the 
constraint $b = 1$, which is equivalent to the assumption $\gamma = 1$, using Eq.~(\ref{scaling}). However, we have also 
relaxed this constraint by treating $b$ as an additional free parameter in the fit. In all cases, no significant deviations 
from $b = 1$ were found for any of the fits. Nevertheless, it would still be desirable to use larger lattices in 
order to minimize the finite-size effects at smaller values of $\beta$.

\begin{figure}[t]
\includegraphics[width=\columnwidth]{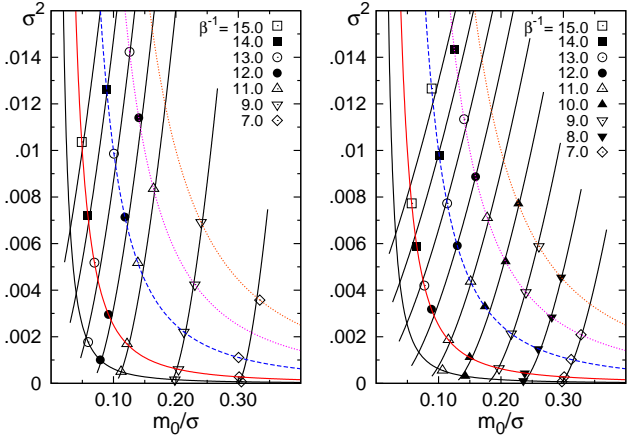}
\caption{\label{fig:Fisher} (Color online) Fisher plot for lattice sizes of $20^2\times20$ (left panel) and $28^2\times28$ 
(right panel). The curved lines connect datapoints of equal $m_0^{}$, from the lower left-hand corner these are $m_0^{} = 
0.0025, 0.005, 0.010, 0.015$ and $0.020$. The vertical lines of equal $\beta$ correspond to the restricted EOS fits shown in 
Fig.~\ref{fig:chiralcondensate1}. The curvature in these lines indicate deviation from the classical critical exponents. At 
the critical coupling, the extrapolation of the lines of equal $\beta$ crosses the origin. Finite-size effects tend to turn 
the lines clockwise.}
\end{figure}

However, it is significant that the present results for both $20^2\times 20$ and $28^2\times 28$ favor values of $\delta 
\sim 2.3$ and $b \sim 1.0$, which strongly disfavors the classical mean-field exponents $\delta = 3$, $\bar\beta = 1/2$. 
Fits using classical exponents tend to become less and less favored when the lattice volume is increased, which is also 
reflected in the ``Fisher plot'' shown in Fig.~\ref{fig:Fisher}. In particular, consistent fits for $\delta$ can be achieved 
using data for $20^2\times 20$ and $28^2\times 28$ if the fit range is restricted to those datapoints where the finite-size 
effects are under reasonable control, as shown in Fig.~\ref{fig:chiralcondensate1}.

It has been argued in Ref.~\cite{Gorbar} that the semimetal-insulator phase transition should present an essential 
singularity, in the sense that the EOS for zero mass in the broken phase would be given by
\bea
\sigma &=& C_0^{} \exp \left (-\frac{C_1^{}}{\sqrt{\beta_c - \beta}}  \right ),
\eea
with $C_0^{},C_1^{}$ constants. This expression has vanishing derivatives to all orders at the critical point, and is said 
to be characterized by Miransky scaling \cite{Gorbar}. The critical exponents corresponding to such a transition 
are $\delta = 1$, $\bar\beta = \infty$ and $\gamma = 1$. This type of transition has sometimes been referred to as a 
Kosterlitz-Thouless transition, even though strictly speaking the latter does not involve spontaneous symmetry breaking. The 
value $\delta = 1$ is apparently ruled out by the considerable dependence of the susceptibility on $m_0^{}$ even for large 
values of $\beta$, which are far from the transition and where the finite-size effects are small. If the value of $\delta$ 
was close to unity, one would observe a susceptibility which is independent of $m_0^{}$ as the critical point is 
approached. While our data does not favor an interpretation in terms of Miransky scaling, a full consideration of this issue is beyond the scope of the present paper.

%


\subsection{Finite-size effects}

\begin{figure}[b]
\includegraphics[width=\columnwidth]{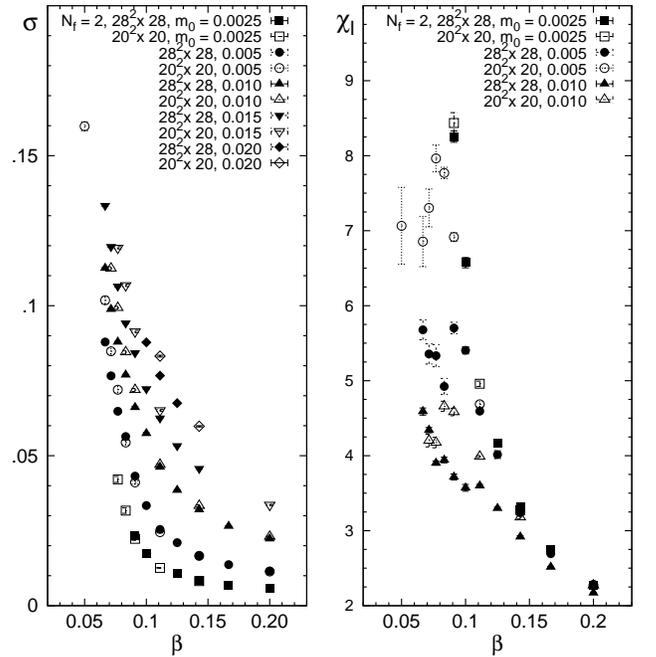}
\caption{\label{fig:volume_effects} Chiral condensate (left panel) and susceptibility (right panel), shown here for 
lattices sizes $20^2\times20$ (open symbols) and $28^2\times 28$ (filled symbols). The finite-size effects are typically 
much larger for the susceptibility, and become large for the condensate as well at small $\beta$, particularly in the broken 
phase. Errors shown are purely statistical, and are in most cases comparable to the size of the datapoints.
}
\end{figure}

If a realistic picture of the properties of the semimetal-insulator transition, as exhibited by the low-energy effective
theory considered here, is to be obtained, a proper assessment of the finite-size effects has to be made. In general, the 
lattice volume should ideally be large enough such that all explicit degrees of freedom (represented in this case by $m_0^{}$) as well as any dynamically generated ones (the Goldstone boson associated with spontaneous chiral symmetry 
breaking) can be contained. In order to illustrate the finite-size effects, the chiral condensate and susceptibility have 
been plotted for volumes of $20^2\times 20$ and $28^2\times 28$ in Fig.~\ref{fig:volume_effects}.

As expected from the quite different nature of the low-energy theory of graphene in the spatial and temporal directions, the 
finite-size effects observed in the simulation are also different. Increasing the extent of the temporal dimension leads to 
an increase in the condensate $\sigma$, as would be expected by comparison with QED$_4^{}$ where the finite-size effects 
are dominated by such behavior. The finite-size effects in the temporal dimension grow as 
$m_0^{}$ is decreased, and do not depend strongly on $\beta$. This indicates that the effects are due to distortion of the 
staggered propagator involving the bare mass $m_0^{}$.

On the other hand, increasing the extent $L_x^{}$ of the spatial directions has a quite different effect on the chiral 
condensate. The effect is to lower the value of $\sigma$, which is opposite to the effect of increasing 
$L_t^{}$. The relative change in $\sigma$ also appears to be roughly independent of $m_0^{}$, such that the absolute 
shift is larger for larger values of $m_0^{}$. It is also noteworthy that the finite-size effects in $L_x^{}$ 
are very small in the unbroken phase, as shown in Fig.~\ref{fig:volume_effects}, while they quickly become large with the onset 
of spontaneous chiral symmetry breaking. We therefore conclude that these effects are due to the emergence of a dynamically 
generated Goldstone mode which is highly spatially extended.

For all the results presented here, the extent $L_z^{} = 8$ has been used for the bulk dimension, in which the fermionic 
degrees of freedom do not propagate. Increasing the size of that dimension has apparently no effect whatsoever on the 
results for the chiral condensate and susceptibility, as demonstrated by a comparison between results on a $14^2\times 14$ 
lattice with $L_z^{} = 8$ and $L_z^{} = 14$. The results for all observables in question are compatible within statistical 
uncertainties, and binning of the data for $\sigma$ into a histogram plot also shows no perceptible differences between the 
two event distributions. Apparently, restricting the bulk dimension to $L_z^{} = 8$ has no significant effect 
on the accuracy of our results, although an increased $L_z^{}$ can be accommodated if necessary, as this has little 
effect on the total computational cost. Such a result is nevertheless somewhat intuitive, as the fermions do not propagate 
in the bulk, and thus should be mostly insensitive to the presence of a boundary in that dimension. However, it should still 
be noted that~\cite{Gockeleretal1,Gockeleretal2,Gockeleretal3} in the context of QED$_4^{}$ the main effect of the boundary 
is to introduce a constant background component into the gauge fields. In other words, at finite volume the results can be 
well described in terms of a renormalized staggered lattice propagator, augmented by a constant background field that may 
vary from one configuration to the next.

In addition to shifting the calculated values of the condensate, finite-size effects may also influence the distribution of 
the measured MC samples. We have observed that for small lattice volumes, the simulation exhibits a tendency to jump between 
two different states, akin to the effect noted in the QCD simulations of Ref.~\cite{Bitaretal}. This effect appears to be 
strongest in the quenched case, and weakens as more fermion flavors are added. The area of parameter space most affected is 
just above $\beta_c^{}$, where the Coulomb interaction is not yet quite strong enough to break the chiral symmetry, and 
$\sigma$ is strongly fluctuating. As this first-order feature also tends to disappear with increasing decorrelation and 
decreasing finite-size effects, we attribute it to a combination of these factors. This is in line with 
Ref.~\cite{Bitaretal}, where attempts to fit the event distribution with two Gaussians did not turn out satisfactorily.


\section{Tests and cross-checks}
\label{sec:tests}

In this section, we briefly describe the various tests performed in order to validate our simulations. Using the formalism 
described in Sec.~\ref{sec:lattice}, we extended our code to perform simulations of QED in 2+1 dimensions (QED$_3^{}$), and 
compared our results with those from Ref.~\cite{Kogutetal}. In this case the differences with graphene are that the gauge 
field lives in one less spatial dimension, and that all the components of the gauge field are dynamical, since Lorentz 
invariance is respected.

\begin{figure}[b]
\includegraphics[width=0.9\columnwidth]{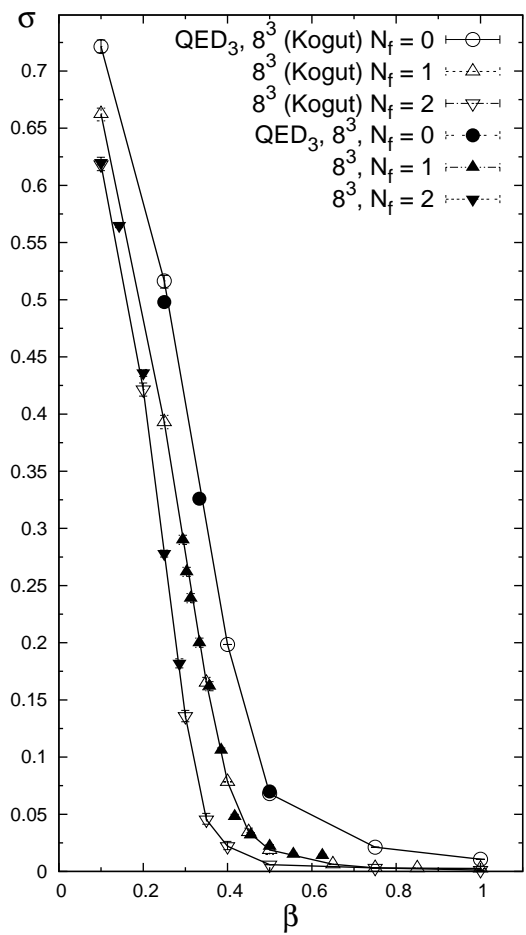}
\caption{\label{fig:QED3_comparison} Data of Ref.~\cite{Kogutetal} compared with our implementation of the QED$_3^{}$ 
simulation. The filled datapoints are our results, whereas the empty ones denote the results of Ref.~\cite{Kogutetal}. The 
lines connecting the datapoints of Ref.~\cite{Kogutetal} are intended as a guide to the eye.} 
\end{figure}

We have also developed another test based on QED in 3+1 dimensions (QED$_4^{}$), which we compared with the results of 
Ref.~\cite{Gockeleretal1,Gockeleretal2,Gockeleretal3}. In this case the differences with graphene affect the fermion field, 
which lives in one more dimension. As in the previous case all the components of the gauge field contribute, as the theory 
is Lorentz invariant. Our lattice Monte Carlo implementation has satisfactorily passed all of the abovementioned tests. A 
comparison between our results for QED$_3^{}$ and those of Ref.~\cite{Kogutetal} is shown in Fig.~\ref{fig:QED3_comparison}.

In addition to these major checks, the following usually overlooked ones were also performed: explicit verification using 
a Computer Algebra System~(CAS) of the correct structure of the staggered fermion operator, invariance of the action and 
the observables under gauge transformations, and reversibility of the HMC algorithm within each MD evolution. We finally 
note that the computing time required by the present calculations is $\sim 10^5_{}$ CPU-hours, which is in line with an 
estimate given by Hands and Strouthos in Ref.~\cite{HandsStrouthos}. Allocations of this size are routinely available 
at various supercomputing centers.


\section{Observation of the Transition}
\label{sec:discussion}

The experimental detection of excitonic instabilities in graphene depends on the size of the induced gap 
$\Delta$. Unfortunately, computing $\Delta$ in absolute units requires knowledge of a suitable dimensionful observable 
(other than $\Delta$ itself) to calibrate the calculation. To our knowledge such a quantity is not yet available. In 
Ref.~\cite{Leal:2003sg}, the excitonic gap was estimated within a gap-equation approach by assuming a value of the cutoff of 
the order of the inverse lattice constant of the graphene honeycomb lattice. In that study, the gap was found to be of the 
order of a few tens of~K. However, such a procedure only constitutes an order-of-magnitude estimate. As in our approach, the 
size of the gap can be determined in absolute units only after calibration of the calculation using a dimensionful 
observable.

Another issue of significance from the experimental point of view is the effect of impurities and lattice defects. These 
were investigated in Ref.~\cite{defects1}, and they were found to have a substantial impact on the low-energy excitations in 
graphene. Also, Ref.~\cite{defects2} has studied the stability of the excitonic insulating phase in the presence of 
impurities, lattice defects and thermal fluctuations, and concluded that all of these effects tend to suppress the excitonic 
instability. Clearly, the experimental demonstration of the semimetal-insulator transition in graphene will be challenging 
from the point of view of sample quality.

As the mere presence of a substrate will likely eliminate the insulating phase due to screening of the Coulomb interaction, 
the most favorable experimental setup would involve samples of suspended graphene. Fortunately, this may also serve to 
eliminate most of the abovementioned concerns. Indeed, it has recently been found in Ref.~\cite{SuspExp} that in order to 
access the intrinsic electronic properties of graphene, thorough current-annealing of suspended samples is necessary. The 
annealed samples were found to exhibit a greatly improved carrier mobility, far in excess of the values reported for 
conventional samples on a substrate. Also, the demonstration of 
Shubnikov-deHaas~(SdH) oscillations suggests that the mean free path in current state-of-the-art suspended graphene is 
comparable to presently achievable sample dimensions of a few $\mu$m. Thus, graphene samples of sufficient quality to 
demonstrate the excitonic instability will likely be available in the near future.

To summarize, our work in Ref.~\cite{DrutLahde1} indicates that the excitonic insulating effect in graphene is unlikely to 
be observed unless the graphene sheet is freely suspended, such that the Coulomb interaction is not screened by the 
dielectric substrate. Further, the experimental work in Ref.~\cite{SuspExp} has demonstrated that the elimination of 
impurities and defects is necessary in order to access the intrinsic electronic properties of graphene. As both of these 
conditions can nowadays be fulfilled by experiment, we hope that the appearance of the excitonic gap will be demonstrated 
in the near future.


\section{Conclusions}
\label{sec:conclusions}

We have described the low-energy effective theory of graphene, its gauge and global symmetries, and shown how a 
discretized lattice formulation can be constructed such that it contains the correct number of degrees of freedom and 
partially retains chiral invariance at finite lattice spacing. We have also explained in detail the numerical methods 
employed to perform lattice Monte Carlo simulations of the discretized theory, focusing on the determination of the location 
and properties of the semimetal-insulator phase transition.

On the theoretical side, we conclude that our extended analysis is consistent with the findings of 
Ref.~\cite{DrutLahde1}, which predict that suspended graphene should possess an excitonic gap in the band structure. We have 
now, using the HMC algorithm, extended the results of Ref.~\cite{DrutLahde1} to much larger lattice volumes, as well as 
smaller fermion masses. While the scenario 
first reported in Ref.~\cite{DrutLahde1} is confirmed by the present results, the larger lattices used also provide 
tantalizing hints that the phase transition is not of infinite order, as predicted in Ref.~\cite{Gorbar}, nor is it likely 
to be described by classical critical exponents. In order to achieve a precise determination of the critical exponents it is 
necessary to perform simulations at much larger lattices, potentially as large as $48^2\times 48$. We are currently 
exploring the feasibility of such simulations by benchmarking our code on a $36^2\times36$ lattice.

An accurate determination of the critical coupling and the critical exponents will provide a solid understanding of the 
universality class of this transition, as well as another piece of experimentally verifiable information on the 
electronic properties of graphene.


\begin{acknowledgments} 

We thank A.~Andreev, W.~Detmold, M.~M.~Forbes, R.~J.~ Furnstahl, D.~Gazit, Y.~Nishida and D.~T.~Son for instructive 
discussions and encouragement, and A.~Bulgac and M.~Savage for providing computing time at the University of Washington. 
We also acknowledge helpful comments by V.~Miransky and Ph.~de~Forcrand. This work was supported in part by an 
allocation of computing time from the Ohio Supercomputer Center, by the National Science Foundation under Grant 
No.~PHY--0653312 and U.S. DOE Grants DE-FC02-07ER41457 (UNEDF SciDAC Collaboration) and DE-FG-02-97ER41014.

\end{acknowledgments}



\begin{thebibliography}{99}

\bibitem{GeimNovoselov}
  K.~S.~Novoselov,
  Science {\bf 306}, 666 (2004);
  K.~S.~Novoselov, D.~Jiang, T.~Booth, V.~V.~Khotkevich, S.~M.~Morozov, A.~K.~Geim,
  Proc.\ Natl.\ Acad.\ Sci.\ U.S.A {\bf 102}, 10451 (2005);
  Nature~(London) {\bf 438}, 197 (2005);
  A.~K.~Geim, K.~S.~Novoselov, 
  Nat.\ Mat.\ {\bf 6}, 183 (2007).

\bibitem{CastroNetoetal}	
  A.~H.~Castro Neto, F.~Guinea, N.~M.~R.~Peres, K.~S.~Novoselov, A.~K.~Geim,
  Rev.\ Mod.\ Phys. {\bf 81}, 109 (2009).

\bibitem{SuspExp}
  K.~I.~Bolotin, K.~J.~Sikes, Z.~Jiang, M.~Klima, G.~Fudenberg, J.~Hone, P.~Kim, H.~L.~Stormer, Solid State Commun. {\bf 
146}, 351 (2008); K.~I.~Bolotin, K.~J.~Sikes, J.~Hone, H.~L.~Stormer, P.~Kim, Phys. Rev. Lett. {\bf 101}, 096802 
(2008); V. Crespi, Physics {\bf 1}, 15 (2008);
  J.~C.~Meyer, A.~K.~Geim, M.~I.~Katsnelson, K.~S.~Novoselov, T.~J.~Booth, S.~Roth, Nature {\bf 446}, 60 (2007);
  J.~S.~Bunch, A.~M.~van~der~Zande, S.~S.~Verbridge, I.~W.~Frank, D.~M.~Tanenbaum, J.~M.~Parpia, H.~G.~Craighead, 
P.~L.~McEuen, Science {\bf 315}, 490 (2007).

\bibitem{Semenoff} 
  G.~W.~Semenoff, 
  Phys.\ Rev.\ Lett. {\bf 53}, 2449 (1984).

\bibitem{DrutLahde1} 
  J.~E.~Drut, T.~A.~L\"ahde,
  Phys.\ Rev.\ Lett. {\bf 102}, 026802 (2009).

\bibitem{Leal:2003sg}
  D.~V.~Khveshchenko, H.~Leal,
  Nucl.\ Phys.\ B {\bf 687}, 323 (2004);
  D.~V.~Khveshchenko,
  Phys. Rev. Lett. {\bf 87}, 246802 (2001).

\bibitem{Gorbar}
  E.~V.~Gorbar, V.~P.~Gusynin, V.~A.~Miransky, I.~A.~Shovkovy,
  Phys.\ Rev.\ B {\bf 66}, 045108 (2002).

\bibitem{Gonzalez} 
  J.~Gonz\'alez, F.~Guinea, M.~A.~H.~Vozmediano,
  Nucl.\ Phys.\ B {\bf 424}, 595 (1994); 
  Phys. Rev. Lett. {\bf 77}, 3589 (1996); 
  Phys. Rev. B {\bf 59}, 2474(R) (1999).

\bibitem{Herbut}
  I.~F.~Herbut, 
  Phys. Rev. Lett. {\bf 97}, 146401 (2006).
  
\bibitem{HandsStrouthos}
  S.~J.~Hands, C.~G.~Strouthos, 
  Phys. Rev. B {\bf 78}, 165423 (2008).

\bibitem{Wallace}
  P.~R.~Wallace, Phys. Rev. {\bf 71}, 622 (1947).
 
\bibitem{Reich}
  S.~Reich, J.~Maultzsch, C.~Thomsen, P.~Ordej\'on,
  Phys. Rev. B {\bf 66}, 035412 (2002).

\bibitem{Son} 
  D.~T.~Son,
  Phys. Rev. B {\bf 75}, 235423 (2007).

\bibitem{Khveshchenko2}  
  D.~V.~Khveshchenko,
  J. Phys.: Condens. Matter {\bf 21}, 075303 (2009).

\bibitem{Rothe} 
 H.~Rothe, 
 ``Lattice Gauge Theories'', $3^\mathrm{rd}$ edition, World Scientific (2005).

\bibitem{compact}
  K.~Farakos, G.~Koutsoumbas,
  Phys.\ Lett.\ B {\bf 178}, 260 (1986);
 J.~B.~Kogut, C.~G.~Strouthos,
 Phys. Rev. D {\bf 71}, 094012 (2005).

\bibitem{NielsenNinomiya}
  H.~B.~Nielsen, M.~Ninomiya,
  Nucl.\ Phys.\  B {\bf 185}, 20 (1981)
  [Erratum-ibid.\  B {\bf 195}, 541 (1982)];
  {\it ibid.} {\bf 193}, 173 (1981).

\bibitem{Kogut-Susskind} 
  J.~Kogut, L.~Susskind, Phys. Rev. D {\bf 11}, 395 (1975);
  L.~Susskind, {\it ibid.} {\bf16}, 3031 (1977);
  H.~Kluberg-Stern, Nucl. Phys. B {\bf 220}, 447 (1983).

\bibitem{KawamotoSmit}
  N.~Kawamoto, J.~Smit,
  Nucl.\ Phys.\ B, {\bf 192}, 100 (1981).

\bibitem{BurdenBurkitt} 
  C.~Burden, A.~N.~Burkitt, 
  Eur.\ Phys.\ Lett. {\bf 3}, 545 (1987).

\bibitem{rooting}
  M.~Creutz,
  Phys.\ Lett.\ B {\bf 649}, 230 (2007). 

\bibitem{PARDISO}
O.~Schenk, K.~G\"artner, 
Future Generation Computer Systems, {\bf 20}, 475 (2004);
Elec.\ Trans.\ Numer.\ Anal. {\bf 23}, 158 (2006). 

\bibitem{BiCGStab}
H.~A.~van~der~Vorst,
SIAM (Soc. Ind. Appl. Math.) J.\ Sci.\ Stat.\ Comput. {\bf 13}(2), 631 (1992).

\bibitem{Bitaretal}
K.~Bitar, A.~D.~Kennedy, R. Horsley, S. Meyer, P. Rossi,
Nucl. Phys. {\bf 313}, 348 (1989); {\it ibid.} {\bf 313}, 377 (1989).

\bibitem{Metropolis}
N.~Metropolis, A.~W.~Rosenbluth, M.~N.~Rosenbluth, A.~H.~Teller, E.~Teller,
J.~Chem.~Phys.~{\bf 21}, 1087 (1953).

\bibitem{DKPR}
  S.~Duane, A.~D.~Kennedy, B.~J.~Pendleton, D.~Roweth,
  Phys.\ Lett.\  B {\bf 195}, 216 (1987).

\bibitem{Gottlieb}
 S.~Gottlieb, W.~Liu, D.~Toussaint, R.~L.~Renken, R.~L.~Sugar,
 Phys.\ Rev.\ D {\bf 35}, 2531 (1987).

\bibitem{DeGrandDeTar}
 T.~DeGrand, C.~DeTar,
 {\it Lattice Methods for Quantum Chromodynamics},
 (World Scientific, 2006).

\bibitem{jackknife}
  M.~C.~K.~Yang, D.~H.~Robinson,
  ``Understanding and learning science by computer'', Series in Computer Science, Vol.~4,
  World Scientific (1986).

\bibitem{Kogutetal}
  J.~B.~Kogut, E.~Dagotto, A.~Koci{\'c},
  Phys. Rev. Lett. {\bf 60}, 772 (1988);
  E.~Dagotto, J.~B.~Kogut, A.~Koci{\'c}, 
  Phys. Rev. Lett. {\bf 62}, 1083 (1989);
  S.~J.~Hands, J.~B.~Kogut, C.~G.~Strouthos,
  Nucl.\ Phys.\  B {\bf 645}, 321 (2002).
  
\bibitem{Kocic:1992pf}
  A.~Koci{\'c}, J.~B.~Kogut, K.~C.~Wang,
  Nucl.\ Phys.\  B {\bf 398}, 405 (1993).

\bibitem{Gockeleretal1} 
  M.~G\"ockeler, R.~Horsley, E.~Laermann, P.~E.~L.~Rakow, G.~Schierholz, R.~Sommer, U.~J.~Wiese,
  Nucl.\ Phys.\ B {\bf 334}, 527 (1990).
  
\bibitem{Gockeleretal2}   
  M.~G\"ockeler, R.~Horsley, P.~Rakow, G.~Schierholz, R.~Sommer,
  Nucl.\ Phys.\ B {\bf 371}, 713 (1992).
  
\bibitem{Gockeleretal3}
 M.~G\"ockeler, R.~Horsley, V.~Linke, P.~E.~L.~Rakow, G.~Schierholz, H.~St\"uben,
  Nucl.\ Phys.\ B {\bf 487}, 313 (1997).

\bibitem{AliKhan:1995wq}
  A.~A.~Khan,
  Phys.\ Rev.\  D {\bf 53}, 6416 (1996).

\bibitem{defects1}
  V.~M.~Pereira, J.~M.~B.~Lopes dos Santos, A.~H.~Castro Neto,
  Phys.\ Rev.\ B {\bf 77}, 115109 (2008); 
  A.~La~Magna, I.~Deretzis, G.~Forte, R.~Pucci,
  Phys.\ Rev.\ B {\bf 78}, 153405 (2008).
  
\bibitem{defects2}
  G.-Z.~Liu, W.~Li, G.~Cheng,
  [arXiv:0811.4471].

\end{thebibliography}
\end{document}